\PassOptionsToPackage{letterpaper,left=1in,right=1in,top=1in,bottom=1in,
                      includeheadfoot,twoside=false}{geometry}

\documentclass[pdflatex,sn-apa]{sn-jnl}% APA Reference Style
\usepackage{graphicx}%
\usepackage{multirow}%
\usepackage{amsmath,amssymb,amsfonts}%
\usepackage{amsthm}%
\usepackage{mathrsfs}%
\usepackage[title]{appendix}%
\usepackage{xcolor}%
\usepackage{tikz} %Added for 8-step chain image
\usepackage{textcomp}%
\usepackage{manyfoot}%
\usepackage{booktabs}%
\usepackage{booktabs}%
\usepackage{tcolorbox}
\usepackage{csquotes} %Added for Einstein's quote on stellar aberration
\usepackage{enumitem} %Added for having automatic "Step..." in reconstructions
\usepackage{algorithm}%
\usepackage{algorithmicx}%
\usepackage{algpseudocode}%
\usepackage{listings}%
\theoremstyle{thmstyleone}%
\theoremstyle{thmstyletwo}%

\theoremstyle{thmstylethree}%

\begin{document}

\title{How are Scientific Concepts Birthed? \par Typing Rules of Concept Formation in Theoretical Physics Reasoning}


\author*[]{\fnm{Omar} \sur{Aguilar}\texorpdfstring{\textsuperscript{*}}{*}} 

\author[]{\fnm{Anthony} \sur{Aguirre}} %\email{iiauthor@gmail.com}

\affil[]{\orgdiv{Physics Department}, \orgname{University of California, Santa Cruz}, \orgaddress{\street{1156 High Street}, \city{Santa Cruz}, \postcode{95064}, \state{CA}, \country{USA}}}

% \affil[2]{\orgdiv{Department}, \orgname{Organization}, \orgaddress{\street{Street}, \city{City}, \postcode{10587}, \state{State}, \country{Country}}}


\abstract{This work aims to formalize some of the ways scientific concepts are formed in the process of theoretical physics discovery. Since this may at first seem like a task beyond the scope of the exact sciences (natural and formal sciences), we begin by presenting arguments for why scientific concept formation can be formalized. Then, we introduce type theory as a natural and well-suited framework for this formalization. We formalize what we call ``ways of discovering new concepts" including concept distinction, property preservation, and concept change, as cognitive typing rules. Next, we apply these cognitive typing rules to two case studies of conceptual discovery in the history of physics: Einstein’s reasoning leading to the impossibility of frozen waves, and his conceptual path to the relativity of time. In these historical episodes, we recast what a physicist might informally call ``ways of discovering new scientific concepts" as compositional typing rules built from cognitive typing rules—thus formalizing them as scientific discovery mechanisms. Lastly, we computationally model the type-theoretic reconstruction of Einstein's conceptual path to the relativity of time as a program synthesis task.}




\maketitle

\begingroup
\renewcommand\thefootnote{}%
\footnotemark
\footnotetext{*Corresponding author:  \href{mailto:omalagui@ucsc.edu}{omalagui@ucsc.edu}}
\endgroup

\section{Introduction} \label{sec:intro}

\vspace{1em}
\begin{quotation}
\noindent\itshape
``Often the right questions are very simple. They invite us to become surprised about perfectly ordinary things. Things that we have taken for granted.”
 —  \normalfont{Noam Chomsky}
\end{quotation}
\vspace{1em}

What is the scientific method? At first glance this question might seem perfectly ordinary. After all, we are introduced to it in middle school. We are taught that it is the instrument that enables science to be possible. It consists of four steps: observing the natural world, formulating a hypothesis to explain these observations, testing the hypothesis against evidence to determine its validity, and repeating the process as needed. Yet, as soon as one embarks on scientific research in college or graduate school, it becomes apparent that this answer is profoundly incomplete. It leaves a fundamental preceding question unaddressed: How does one arrive at a hypothesis in the first place? or more broadly: 

\begin{center}
    \textit{How does one create a scientific concept?}
\end{center}

Faced with this conundrum, a physicist may respond in one of two broad ways. They may say that the act of creating scientific concepts is subjective and almost mystical, hence inaccessible to rational analysis—and so they simply stop there, because there's nothing left to explain. Or, they may acknowledge that scientific concept formation is at the very least partially subjective, but still try to make sense of it—by reflecting on how they and other scientists create concepts in their natural creative environments, namely conversations and internal monologues, and then finding patterns in those reflections. If they take the second choice, they might recall learning early on that Newton imagined the moon as a falling apple. They might also recall learning in class that one of Einstein’s first steps toward special relativity was his refusal to abandon the constancy of the speed of light. Once equipped with these canonical results in fundamental physics, they may apply their training to frontier questions in the physics of complex systems and ask—while searching for a measure of complexity—``What would an entropy-like measure look like if, instead of always increasing in a closed system, it rose to a peak and then declined?” \citep{aaronson2014coffee, crutchfield1994calculi}. If one abstracts from these memories, a common thread emerges: they all involve the imposition of \textit{constraints}. This, in essence, was the first concrete answer in recorded history to the question of how scientific concepts are generated—offered by the 19th century philosopher and polymath C.S. Peirce.

Peirce proposed that forming scientific concepts is \textit{abduction} \citep{chomsky2006peirce}—the process of imposing constraints on admissible hypotheses to generate plausible, though fallible, explanations\footnote{In support of this, he mentioned that in the history of science, once certain constraints were agreed upon, different scientists without communicating with each other came to the same conclusion \citep{chomsky2006peirce}}. A Peircean notion of a scientific concept follows from this view: it is an entity that satisfies the constraints of the process through which it is discovered. As the 20th century began, however, interest in how scientific concepts are formed—and in defining them in terms of their formation—declined within mainstream philosophy of science, as scientific concepts came to be treated solely as ideal entities waiting to be discovered \citep{Feigl1970}. 

A small resurgence of interest in scientific concept formation began in the second half of the century led by small, scattered groups of philosophers who drew on insights from the history of science \citep{hanson1958patterns, kuhn1962structure} and, by the 1980s, from cognitive science \citep{nersessian1987coghistanalysis, thagard1988computational, gentner1983structure} as well. These combined insights gave rise to a newly established interdisciplinary field dedicated to the study of scientific cognition: the \textit{cognitive science of science} \citep{thagard2012cognitive, magnani2007model, gorman2005scientific}. Like Peirce, researchers in this field conceived scientific concepts in terms of how they were discovered, but this time their definition was much more precise: they treated concepts as entities shaped by the constraints of the knowledge-construction practices scientists use to generate them—such as analogies and thought experiments \citep{nersessian2008creating}.
%\footnote{Although figures like Popper and Kuhn are widely recognized among scientists, they did not explicitly engage with this question}

%multidisciplinary approaches to address the question. Their efforts were deeply shaped by

% Cognitive scientist of science Nancy Nersessian has provided a synthesis of the cognitive and historical aproaches, known as \textit{cognitive-historical analysis}, that emphasizes processes over products. 

%most informed by detailed knowledge of theoretical physics.

%In her study of concept creation in theoretical physics, she shifts the focus from concepts themselves to the problem-solving practices scientists use to generate and change them. 

Perhaps the most detailed engagement between the cognitive science of science and theoretical physics is found in the work of Nancy Nersessian, who developed a method for studying scientific concept formation known as \textit{cognitive-historical analysis} \citep{nersessian1987coghistanalysis}. The historical component of her method treats detailed records of theoretical physics practice—such as notebooks, diaries, correspondence, drafts, and publications—as empirical data and subjects them to philosophical analysis. Meanwhile, the cognitive component treats these practices as traces of cognitive mechanisms—analogy, mental modeling, and thought experiments—that allow us move from specific to general conclusions about the nature of scientific concept formation. In effect, Nersessian conceives scientific concepts as entities that satisfy the constraints imposed by discovery methods used to generate them—such as the \textit{method of physical analogy} employed by Maxwell \citep{NersessianMaxwell2002}. Since discovery methods like physical analogy are systematic procedures with a definite and finite number of steps, Nersessian’s framework gives rise to the first central question of this paper:
% —and since computation formalizes precisely such well-defined, finitely executable procedures—

% By combining the hard realism of historical analysis with the broad explanatory power of cognitive science, this framework offers general yet grounded insights into the nature of concept forming practices in theoretical physics. 
% \textbf{\textcolor{violet}{OAC: Define the definition of concept we will use}} These insights reveal that the knowledge-constructing practices in theoretical physics, which shape and define concepts, follow systematic procedures 

\begin{center}
    \textit{Can concept formation in theoretical physics \\ be formalized?}
\end{center}

% \begin{center}
%     \textit{Can concept formation in theoretical physics \\ be formalized computationally?}
% \end{center}

To address this question, one may be inclined to explore machine learning approaches -- such as graph neural networks \citep{battaglia2018relational}, symbolic regression \citep{udrescu2020symbolic}, and transformers \citep{lalande2023transformer} -- that have successfully automated other aspects of scientific discovery, with the aim of adapting these techniques to model scientific concept formation. However, as discussed in \citep{battledey2024AI} and  Section~\ref{sec:relevant_work}, these methods still face the challenge of  modeling conceptual development, as they were built to solve predefined problems rather than define new ones. This suggests that we should take a step back and further clarify scientific concept formation and scientific concepts—by clarifying what has so far been left vague: what we mean by constraints.

As part of this effort, physicists might recall that when searching for new scientific concepts, they often use the word ``type" or ``kind" informally to describe what they are looking for. For instance, when trying to define a thermodynamic entropy for non-equilibrium systems \citep{safranek2021observational}, they might ask, ``What type of entropy would count as thermodynamic, yet increase continuously rather than abruptly as the system transitions from one equilibrium configuration (a closed box) to another (an open box after a long time)?" \citep{safranek2020classical}. In some cases, physicists even explicitly use the words ``type" or ``kind" in a colloquial sense when describing their discovery process in papers—just as Maxwell did when he first conceived of the magnetic field as a ``type" of stress composed of pressures and tensions \citep{maxwell1865lines}. 

These \textit{intuitive types} are not necessarily physically correct claims, but rather common-sense constraints that help narrow the space of possible concepts moving forward. Their intuitiveness stems from the fact that they are used most frequently in everyday human reasoning. For example, one way intuitive types show up in everyday reasoning is in how people interpret the question, ``What time is it?" Most would respond with a time, like ``2 PM" even if it’s incorrect. But they wouldn’t answer, ``A burrito with pinto beans"—unless as a joke—because that response doesn’t match the expected ``type" of answer, which is time. 

% For example, combining the intuitive types of meat and bread by placing meat between two slices of bread may lead one, like John Montagu, the 4th Earl of Sandwich, to the concept of a sandwich, which encompasses both the types meat and bread.

Felix Sosa has recently characterized these intuitive types as cognitive constraints that impose structure on concepts, enabling people to generate reasonable—though not necessarily correct—answers to novel questions \citep{sosa2022type}. Furthermore, he has proposed a natural formalization of these types as \textit{theoretical types} which are the building blocks of \textit{type theory}\footnote{Type theory feels natural because it captures the core of intuitive types—basic distinctions between kinds like numbers and places—without adding unnecessary structure to what those distinctions mean. Its compositional rules reflect how we build complex intuitive types in language, and its constraints filter out incompatible ones before conscious reasoning begins.}. This makes one wonder: if intuitive types used in everyday thought—like ``bread” or ``dog”—are slippery yet still conceivable as theoretical types, what about those used in scientific reasoning, such as ``has 3 distinct components” which are already well-defined? This brings us to the second and more narrowly focused central question of this paper:

\begin{center}
    \textit{Can concept formation in theoretical physics be formalized \\ using type theory? }
\end{center}

% This suggests that we might try going the other way around—first seeking computational approaches that have shown success at modeling concept formation in other cognitive domains, such as concept acquisition in children \textbf{\textcolor{violet}{OAC: Cite}} and common-sense learning \textbf{\textcolor{violet}{OAC: Cite}}, and then extending them to capture scientific concept formation. 

Our answer is yes, and in this paper we set out to substantiate this claim through a type-theoretic formalization of scientific concept formation driven by detailed reconstructions of scientific reasoning from the history of physics. In this framework, intuitive types of scientific concepts are formalized as types and different ``ways of discovering scientific concepts" as typing rules—thereby recasting them as \textit{discovery mechanisms}. Moreover, scientific concepts are formalized both as \textit{terms} and as the results of \textit{typing rules}—thus defining them by how they are discovered, in the spirit of Peirce and cognitive science of science. 
% Thus, in this paper, scientific concepts will be regarded as satisfying types imposed by typing rules used to discover them. 

Lastly, for one familiar with programming—where types like \texttt{true}\footnote{Type names are set in monospaced font to emphasize their formal role and set them apart from the surrounding text. The same convention is used for discovery mechanism names to highlight their central place in the formalism.} and \texttt{int} are standard—it is natural to wonder whether type theory can be viewed through a computational lens. This is indeed the case, as a type system can be understood as a mathematical formalization of a programming language \citep{cardelli1996typesystems}, where terms correspond to programs\footnote{Typically, functions from the lambda calculus, a compact formal language of functions, but also computer programs (e.g., in Python).} and types specify how those programs may be used. This perspective allows us to recast scientific discovery as  \textit{program synthesis}: the inference of programs from data and constraints \citep{rule2020child}. To illustrate this idea, we will carry out type-theoretic reconstructions of two examples of concept formation from the history of physics, and computationally implement one of them using program synthesis.

\section{Overview}

In summary, there are two main reasons to consider type theory as a mathematical language for formalizing concept formation in theoretical physics:

\begin{enumerate}
    \item Theoretical types provide a natural formalization of intuitive types, which theoretical physicists rely on when creating new concepts
    \item Intuitive types used in theoretical physics are often more precisely defined than those used in everyday reasoning, which makes the former much easier to formalize than the latter. For example, an intuitive type like “a force with a wave-like direction” has more sharply defined properties than intuitive types like toy, dog, or sweet, which remain comparatively vague.
\end{enumerate}

This paper aims to provide a type-theoretic formalization of some ways in which scientific concepts are formed in theoretical physics, based on reasoning patterns reconstructed from the history of theoretical physics. To this end, the paper undertakes the following recastings:

\begin{enumerate}
    \item Intuitive types are formalized as theoretical types.
    
    \item ``Ways of discovering new concepts''—such as concept distinction, property preservation and concept change—are formalized as elementary typing rules, which we call cognitive typing rules.
    
    \item ``Ways of discovering new scientific concepts” are formalized as compositional typing rules that integrate cognitive typing rules with functional and algebraic operations. In doing so, these ``ways" are recasted as discovery mechanisms.
    
    \item Scientific concepts are formalized both as terms and as the outcomes of discovery mechanisms.
\end{enumerate}

The formalism we develop is applied to a paradigmatic conceptual breakthrough in the history of physics: Einstein’s special relativity. Specifically, we examine two key episodes in Einstein’s reasoning that paved the way to his discovery. Namely, the conceptual paths that led him to reject the possibility of frozen waves and to recognize the relativity of time. 
% In Maxwell’s case, we analyze the reasoning that led him to recast Faraday’s lines—from a medium of stresses to a surface of vortices.

Einstein’s reasoning leading to the impossibility of frozen waves is formalized as a \texttt{distinction} discovery mechanism, which combines a distinction typing rule with functional and algebraic operations. Moreover, Einstein's reasoning leading to the relativity of time is formalized as a discovery mechanism that combines property preservation and concept change typing rules with functional and algebraic operations. At a global level, this discovery mechanism reverses the roles of assumption and conclusion, which leads us to name it the \texttt{assumption}–\texttt{conclusion switch} discovery mechanism. Lastly, we use typed-program synthesis to simulate Einstein's conceptual path to the relativity of time.

Although many studies have modeled the discovery of symbolic equations \citep{udrescu2020symbolic}, scalars \citep{Zhang2018DeepPotentialMD}, vectors \citep{Schutt2018SchNet}, and probability distributions in physics \citep{Noe2019BoltzmannGensDists}, this paper is, to our knowledge, the first to formalize and model concept discovery in theoretical physics. By integrating insights from cognitive science, computer science, and the history of physics, this study offers a novel perspective that frames concept discovery in theoretical physics as a structured process, rather than an inscrutable one. This approach holds that by taking seriously the history and philosophy of physics as blueprints for the reasoning patterns behind theoretical discoveries—and by formalizing those patterns using type theory and program synthesis—we can begin to address the question of how scientific concepts are formed, scientifically.

The rest of the paper is organized as follows. Section~\ref{sec:relevant_work} compares our formalism to past and current AI approaches to scientific discovery and complex systems frameworks. Section~\ref{sec:background} provides an overview of the type theory background relevant to our goals, emphasizing intuition over rigor to ensure accessibility to researchers from diverse fields such as physics, artificial intelligence, cognitive science, psychology, and philosophy. Section~\ref{sec:methods} presents the cognitive typing rules that underpin our formal approach. In Section~\ref{sec:results}, we apply the type theory background and cognitive typing rules developed in the previous sections to two historical episodes in theoretical physics: Einstein’s reasoning leading to the impossibility of frozen waves and his conceptual path to the relativity of time. These applications lead to the identification of two discovery mechanisms—\texttt{distinction} and \texttt{assumption}–\texttt{conclusion switch}—all of which take the form of compositional typing rules constructed from cognitive typing rules and functional and algebraic operations. Section~\ref{sec:program-induction} computationally models the type-theoretic reconstruction of Einstein's conceptual path to the relativity of time as a program synthesis task. Finally, Section~\ref{sec:conclusion} summarizes the paper, reflects on its broader implications, and outlines directions for future research.

\section{Related work} \label{sec:relevant_work}

The idea of computationally modeling scientific discovery—or using AI for science—began with Herbert Simon and Pat Langley's invention of \textit{symbolic regression}, the computational inference of equations from data \citep{simon1973logic, simon1981scientific}. They implemented it using heuristics inspired by the history of early classical physics, and in the following decades, subsequent implementations—such as those using delay-coordinate embedding \citep{packard1980geometry} and genetic algorithms \citep{schmidt2009distilling}—vastly improved its scalability. More recently, deep learning has not only enhanced the efficiency and scale of symbolic regression \citep{udrescu2020ai}, but also enabled the discovery and rediscovery of other physically meaningful mathematical objects. Among these objects one finds scalars like band gap energies that signal new topological insulators \citep{claussen2020detection}; probability distributions like the Bernoulli model used to confirm Kepler-90i as a real exoplanet \citep{shallue2018identifying}; and operators such as those defining the Navier–Stokes equations \citep{li2020fourier}. As a result, deep learning has become the dominant paradigm in AI for science. 

Yet despite the success of both early and modern AI for science approaches in computationally modeling key aspects of the scientific discovery process, they still face a common challenge: providing a computational account of its conceptual dimension. Their difficulty in doing so stems from framing discovery solely as the optimization of solutions to well-defined questions, rather than also including the generation of new questions—a process that requires conceptual innovation. This distinction between solving well-defined problems and generating new ones was recently examined by Battleday and Gershman \citep{battledey2024AI}, who describe them as the “easy” and “hard” problems of AI for science.

In their recent work, they point out that, at a high level, formulating a new scientific problem is akin to specifying both the domain—identifying the relevant phenomena—and the theoretical constraints—determining the properties that a theory explaining these phenomena must satisfy. Moreover, they note that these specifications involve not only objective aspects but also subjective ones, such as taste and style. They further note that the interaction of these objective and subjective aspects is the central object of study in the cognitive science of science, and therefore, any computational account of scientific question formation should begin with insights from this field. In this context, the present work represents a step in that direction. Moreover, this endeavor can be regarded as a fine-grained complement to coarser-grained, multi-agent computational models of scientific discovery that capture how research communities select strategies for scientific discovery and offer principled guidance on which strategies are most effective \citep{Dubova2022Against}.

At this point, one may wonder: what alternative AI approaches—not informed by the cognitive science of science—are currently being pursued to address the hard problem of AI for science? Battleday and Gershman point out that the only existing candidate is the AI Scientist \citep{lu2024aiscientist}, a system that uses large language models, prompt engineering, and coding assistants to generate scientific questions in natural language. Its outputs are often generated by splitting variables, modifying processing pathways, or adding new components to the model. While the AI Scientist generates human-interpretable outputs and, if sufficiently scalable, may contribute to the generation of new scientific hypotheses or concepts, the procedure by which these outputs are produced still functions largely as a black box \citep{mengaldo2024explain}, providing no formal justification for how the results are derived. As a result, this could increase the risk of overlooking empirically correct scientific concepts that appear unreasonable, or of accepting seemingly reasonable ones that rest on flawed assumptions.

Moreover, although the AI Scientist produces scientific concepts in natural language, it does not currently provide a formal representation of them \citep{wolfram2024can}. For one seeking to formalize scientific concept formation, this makes it unclear how to define scientific concepts precisely and establish exact relationships among them. As a result, one may struggle to refine scientific concepts in a way that reliably generates reasonable, yet fallible, novel concepts or that effectively challenges deeply held assumptions. Furthermore, because the output is solely in natural language and not explicitly restricted to properties relevant in a given context, it is open to multiple interpretations—that is, it is \textit{overloaded with meaning}—which can lead to ambiguous concepts \citep{liu2023ambiguity}. In sum, while the AI Scientist generates promising natural language outputs that could give rise to novel scientific concepts, it does not yet offer a formal account of how such concepts are formed—either in the processes it follows or in the outputs it produces. To address these challenges, we examine how a type-theoretic approach to scientific concept formation might help overcome them.

To begin with, a type-theoretic framework is a formal system built on well-defined objects of study (types) and procedures (typing rules), thereby making the process from input to output explainable rather than a black box. In this system, concepts are formalized as both terms and the outcomes of typing rules, and the concepts' relevant properties are represented as types. This formalism makes it possible to relate concepts objectively—either by assigning them the same type, or by applying rules that govern how concepts of the same or different types interact. Furthermore, the problem of meaning overload can be naturally avoided by the fact that each typing assertion is made relative to a set of assumptions, known as the context. This does not compromise type theory's \textit{syntactic expressiveness}, meaning its capacity to represent concepts in varied syntactic forms (e.g., equations, hierarchical relationships, or visual constraints).  In addition, both the assignment of a type to a term and the typing rules themselves act as constraints\footnote{Constraints are chosen over raw data, since the former are the input that guides the conceptual thinker and prompted the kind of reasoning Maxwell used to discover his electrodynamics laws. He was not trying to fit an equation to the positions of the iron filings around a magnet, as depicted in Faraday’s magnetic iron filings diagram. Instead, he developed a theory that captured the qualitative constraints revealed by the diagram \citep{maxwell1865lines}.} that guide the reasoning process. Lastly, although type theory is an abstract formalism, its syntactic expressiveness allows both its terms and types to retain the concrete features that give scientific concepts their meaning. 

\section{Type theory} \label{sec:background}

Type theory is a branch of mathematical logic that formalizes the classification of mathematical objects, known as terms, by assigning them classifiers called types, which constrain how terms can be used and combined \citep{pierce2002types}. Originally developed to avoid logical paradoxes in set theory, it now may be regarded as a foundation for the construction of mathematics \citep{russell1903principles} and as a formalization of programming languages \citep{pierce2002types}. In the latter role, it has been used to automate basic algebraic problem solving \citep{poesia2023peano}, as well as the formulation and verification of mathematical proofs and conjectures \citep{andrews1996tps}. 

We now aim to extend the use of type theory to model concept formation in theoretical physics reasoning. To this end, we begin by introducing type theory with an emphasis on intuition over rigor, illustrating the key ideas relevant to our purpose through examples drawn from physics. For simplicity, the type-theoretic examples in this section involve only terms that represent fully established physical concepts—those that are clearly physically interpretable and recognized as valid within currently accepted theories. In contrast, examples involving provisional concepts—constructs that merely satisfy certain physical constraints and aid reasoning, but may not be fully physical or valid within any accepted theory—will be presented in Section~\ref{sec:results}.

\subsection{Expressions, terms and types}

In a type system, an \textit{expression} is any sequence of symbols allowed by the system’s typing judgments and typing rules \citep{sep-type-theory-church}. Common-sense examples of this are ``Colorless green ideas sleep furiously” \citep{chomsky1957syntactic} and ``The square root of a triangle equals banana” which are grammatically well-formed but lack coherent, common-sense meaning. Physics examples such as $ \text{Charge} + \text{Temperature} $ or $ \text{Force} = \frac{\text{Meters}}{\text{Seconds}^2} $  are syntactically well-formed in the sense that they resemble mathematical equations, but they are physically meaningless—here because they violate dimensional consistency or because they define a physical quantity solely in terms of units.

A \textit{term} is a specific kind of expression that carries semantic content—that is, it “is or does something” in the type system \citep{sep-type-theory-church}. More precisely, a term is an expression that can be assigned a type in the type system. Common-sense examples of terms include ``a red apple”, ``a wooden chair" or a ``barking dog"—each referring to familiar and recognizable entities. In physics, examples of terms include variables, constants, and function applications such as a particle’s velocity $v$, the speed constant $c$, and the Galilean acceleration equation $\frac{v}{t}$.

A \textit{type} classifies terms and determines their valid uses\footnote{In this paper’s type system, a term may have more than one type.}. This classification is made by asserting that a term $t$ has type $A$, denoted as $t : A$, and the determination of valid use is carried out by judgments and typing rules. Common-sense examples include assigning the term ``a red apple” the type \texttt{fruit}, or ``a wooden chair” the type \texttt{furniture}—reflecting how we often understand things in terms of what kind they are. In physics, this corresponds to assigning the speed constant $c$ the type \texttt{physical-constant}, and the Galilean acceleration expression $v/t$ the type \texttt{kinematic-quantity}, where both the assignment of types to terms and the use of those terms are defined by the system’s typing rules.

\subsection{Contexts}

A \textit{context} $\Gamma$ is a collection of assumptions under which an assertion can be made. For a common-sense example, making a cake occurs under a kitchen context, $\Gamma_{\text{kitchen}}$, which assumes flour, sugar, and an oven are available. Likewise, doing laundry takes place in a laundry‐preparation context, $\Gamma_{\text{laundry prep}}$, which requires sorted clothes, detergent, and an empty washing machine. In physics, contexts encode the theoretical assumptions. For instance, in the Newtonian physics context $\Gamma_{\text{Newton}}$, we assume Newton’s second and third laws; in the classical mechanics (CM) context  $\Gamma_{\text{CM}}$, we adopt a classical mechanical framework; and in the non-relativistic context $\Gamma_{\text{non-rel}}$, we restrict all objects to speeds far below the speed of light.

\subsection{Typing jugements}

A typing judgment asserts that, under a context $\Gamma$, a term $t$ has type $A$, and is denoted as
$$\Gamma \vdash t : A$$

For a common-sense example, consider a coin in two everyday contexts. In the arcade context $\Gamma_{\text{arcade}}$--where coins serve to activate an arcade machine--we assign coins the type \texttt{activator}, while in the wishing-well context $\Gamma_{\text{wishing-well}}$--where coins symbolize chance or hope--we assign them the type \texttt{luck}. Formally, we write
$$
\Gamma_{\text{arcade}} \vdash \text{coin} : \texttt{activator} \quad \text{and} \quad
\Gamma_{\text{wishing-well}} \vdash \text{coin} : \texttt{luck}.
$$
In physics, within the low speed classical mechanics context $\Gamma_{v\ll c,\text{CM}}$, kinetic energy of type \texttt{energy} takes the form $KE = \frac{1}{2}mv^2$. Formally, we express this by the following typing judgment:
$$
\Gamma_{\text{CM}} \vdash KE = \tfrac{1}{2} m v^2 : \texttt{energy}.
$$

Now consider the physical quantity angular momentum, of type \texttt{momentum}, which takes different mathematical forms in two distinct theoretical physics contexts. First, within the classical mechanical (CM) context $\Gamma_{\text{CM}}$, the angular momentum $L$ of a rotating system is given by $L = mvr$. However, within the quantum mechanical (QM) context $\Gamma_{\text{QM}}$, angular momentum is quantized and takes discrete values given by the equation $L = \sqrt{\ell(\ell + 1)}\,\hbar$, where $\ell$ can be any non-negative integer. Formally, these judgments are denoted as $$\Gamma_{\text{CM}} \vdash L = m v r : \texttt{momentum} \quad \text{and} \quad \Gamma_{\text{QM}} \vdash L = \sqrt{\ell(\ell + 1)}\hbar : \texttt{momentum}$$

\subsection{Typing rules}

In order to specify how a typing judgment is derived, it is often defined \textit{inductively} \citep{harper2016practical}. That is, a typing judgement is defined as a collection of \textit{typing rules}. Each typing rule defines a typing judgment as a conclusion derived from premises that are themselves already established judgments. Schematically, a typing rule is written in the following form\footnote{In a typing rule, the horizontal bar denotes that the typing judgment written beneath is \textit{derived} from the typing judgments above. In contrast, the entailment relation $\vdash$ indicates that the judgment on its right is \textit{derivable} in the context on its left; the assumptions are required, yet their presence alone is not enough to derive the judgment.}:
$$
\frac{\text{premise} \ 1 \quad \cdots \quad \text{premise} \ n}{\text{conclusion}}
$$

Formally, a typing rule asserts that the typing judgment $t : A$ holds under the context $\Gamma$ if each of its premises $t_i : A_i$ holds under the context $\Gamma, \Gamma_i$\footnote{Context extension (comma) forms the context $\Gamma, \Gamma_i$ by appending $\Gamma_i$ to $\Gamma$ and requires that $\Gamma_i$ introduce no duplicate variables. In contrast, we later use context union ($\cup$) to merge two independently built contexts that may share variable names.}, where $\Gamma_i$ is a local extension of $\Gamma$ specific to the $i^{\text{th}}$ premise. This is denoted as follows:
$$
\frac{\Gamma, \Gamma_1 \vdash t_1 : A_1 \quad \ldots \quad \Gamma, \Gamma_n \vdash t_n : A_n}{\Gamma \vdash t : A}
$$

At this point, we offer a few remarks on the nature of typing rules. A rule with no premises holds unconditionally and is therefore called an \textit{axiom}. Moreover, the definition of a typing rule implies that the premises are sufficient to derive the conclusion. Furthermore, different collections of premises can lead to the same conclusion. Thus, even if a conclusion holds, it does not mean that any specific set of premises must have held. Moreover, since $\Gamma$ applies uniformly to all premises and the conclusion, it can be omitted. This yields the \textit{local form} of the inductive definition, as opposed to the \textit{global form} in the immediately preceding displayed rule \citep{harper2016practical}.

To illustrate typing rules using familiar scenarios, consider the following examples. First, suppose that under the kitchen context $\Gamma_{\text{kitchen}}$, the term banana has type \texttt{fruit}. Moreover, under the same context, the term blender has type \texttt{fruit} $\rightarrow$ \texttt{smoothie}, meaning it takes an input of type \texttt{fruit} and produces an output of type \texttt{smoothie}. From these premises, we can infer that the function application term blender(banana) has type \texttt{smoothie}. This is a common-sense example of a standard function application typing rule. For the second example, suppose that under the mood context $\Gamma_{\text{mood}}$, both terms happy and sad have type \texttt{emotion}. From these premises, we can form the pair term (happy, sad) which may represent nostalgia or catharsis, and is assigned the type \texttt{emotion}. This is a common-sense example of a standard pair typing rule that preserves the type of its components. The examples are formalized as follows:

$$
\dfrac{
  \Gamma_{\text{kitchen}} \vdash \text{blender} : \texttt{fruit} \rightarrow \texttt{smoothie} \quad 
  \Gamma_{\text{kitchen}} \vdash \text{banana} : \texttt{fruit}
}{
  \Gamma_{\text{kitchen}} \vdash \text{blender(banana)} : \texttt{smoothie}
}
$$

$$
\dfrac{
  \Gamma_{\text{mood}} \vdash \text{happy} : \texttt{emotion} \quad 
  \Gamma_{\text{mood}} \vdash \text{sad} : \texttt{emotion}
}{
  \Gamma_{\text{mood}} \vdash (\text{happy}, \text{sad}) : \texttt{emotion}
}
$$

The following physics examples  illustrate how typing rules operate—either by directly assigning a type to a basic term or by building the type of a complex expression from the types of its parts. These include a constant rule for the gravitational constant $g$, a function application\footnote{In lambda calculus, function abstraction \textit{defines} a function and is written $\lambda x.t$, meaning ``the function that takes $x$ and returns $t$”. Moreover, function application \textit{uses} a function and is written $t_1 t_2$, meaning “apply the term $t_1$ to the argument term $t_2$.”} rule for uniform linear velocity $v=\frac{x}{t}$ and a function abstraction rule for uniform linear acceleration $a=\frac{v}{t}$, as shown below:
$$\frac{}{\Gamma \vdash g : \texttt{real}}, \quad    \frac{\Gamma \vdash x: \texttt{pos} \quad \Gamma \vdash t : \texttt{time} \quad \Gamma \vdash v : \texttt{pos} \to \texttt{time} \to \texttt{vel}}{\Gamma \vdash v(x, t) : \texttt{vel}}$$
$$\text{and} \quad \frac{\Gamma, \ v : \texttt{vel}, \ t : \texttt{time} \vdash v/t : \texttt{acc}}{\Gamma \vdash \lambda v. \lambda t. v/t : \texttt{vel} \to \texttt{time} \to \texttt{acc}}$$

\subsection{Equality Judgments}

Type theory allows us to relate scientific concepts in diverse ways through distinct notions of equality. Each such notion is expressed by an \textit{equality judgment} comprising three claims: two typing judgments (taken as implicit premises) and an equality assertion between the terms according to the chosen notion. A helpful intuition for distinguishing these notions of equality is captured by Bhavik Mehta's saying: ``Syntactic equality is they look identical, definitional equality is they are the same, propositional equality is they turn out to be the same" \citep{buzzard2024formalising}. Below, we present their formal definitions\footnote{While we use distinct notation here to clarify the different notions of equality, in Sections~\ref{sec:methods} and~\ref{sec:results} the intended notion of equality will be inferred from context}.

\textbf{Syntactic Equality.} This equality holds between two terms $t$ and $t'$ when they are identical in written form. That is, they consist of the same sequence of symbols, ignoring trivial differences such as whitespace or renaming bound variables\footnote{Differing only in the names of bound variables is known as \textit{$\alpha$-equivalence} in lambda calculus}. For a common-sense example, consider the syntactic equality $\text{cat} = \text{cat}$. For a mathematical example, consider the lambda functions $(\lambda y. y + 1)$ and $(\lambda x. x + 1)$. These are syntactically equal as they differ only in the name of the bound variable.

\textbf{Definitional Equality.} This equality generalizes syntactic equality by stating that two terms $t$ and $t'$, both of type $A$, are equal if they are \textit{computationally identical}, meaning they reduce to the same normal form. It is written as: 
$$\Gamma \vdash t \equiv t' : A$$

A common-sense example of definitional equality is the equivalence between the phrases ``The capital of Peru” and ``Lima” since they refer to the same entity. In physics, a clear example is the equivalence between the force terms $\frac{dp}{dt}$ and $m \frac{dv}{dt}$ (assuming mass is constant), which are definitionally equal after expanding the definition of momentum $p$\footnote{via $\delta$-reduction, in the language of lambda calculus}. This can be expressed as $\Gamma \vdash \frac{dp}{dt} \equiv m \frac{dv}{dt} : \texttt{force}$. 

\textbf{Propositional Equality.} This extends definitional equality by stating that two terms $t$ and $t'$ of the same type $A$ are equal if their equivalence can be \textit{constructed} (proven) by a sequence of computational transformations rather than direct reduction. It is expressed as:
$$
\Gamma \vdash t = t' : A.
$$  

A common-sense example of propositional equality is the equivalence between the phrases  ``The package at your door" and  ``the item you ordered online last week" as both refer to the same entity, though recognizing this requires some reasoning. In physics, an example of this is given by the energy terms $\int F \cdot dx$ and $\frac{1}{2} m v^2 - \frac{1}{2} m v_0^2$, which are propositionally equal after integration. This can be written as $\Gamma \vdash \int F \cdot dx = \frac{1}{2} m v^2 - \frac{1}{2} m v_0^2 : \texttt{energy}$.

\textbf{Heterogeneous Equality.} This equality generalizes propositional equality by allowing terms of different types to be equal. It is expressed as:
$$
\Gamma \vdash (t : A) = (t' : B)
$$

A common-sense example of heterogeneous equality is the equivalence between the action of waving your hand when someone walks in and saying ``Hello!”—two expressions of the same intent, even though they are of different kinds. In physics, an example of heterogeneous equality is the equivalence between the gravitational force $F_g = \frac{G M m}{r^2}$ and centripetal force $F_c = F_c = \frac{m v^2}{r}$, which are equal in the case of a circular orbit under gravity. This can be expressed as $\Gamma \vdash F_g : \texttt{grav} = F_c : \texttt{cent}$

\subsection{Special Types}

Below, we present non-elementary types that are necessary for describing scientific concept formation using type theory.

\textbf{Intersection Type.} An intersection type $A \cap B$ is a type that combines types $A$ and $B$, such that any term of type $A \cap B$ has both types $A$ and $B$. A common-sense example is the coin flip, which functions both as an activator in arcades and as a symbol of luck in wishing wells—giving it the intersection type $\texttt{activator} \cap \texttt{luck}$. In physics, a charged pendulum swinging in Earth’s magnetic field is governed by both the laws of mechanics and the laws of electrodynamics—giving it the intersection type $\texttt{mech} \ \cap \ \texttt{elect}$. This can be expressed as $\Gamma \vdash \text{charged-pend} : \texttt{mech} \ \cap \ \texttt{elect}$.

\textbf{Contradiction Type.} The contradiction type $\bot$ is a type for which no term can be constructed unless there is an inconsistency.  Thus, if a derivation results in a term of type $\bot$, this indicates that some of the assumptions used in the derivation are inconsistent. For a common-sense example, consider the statement ``the light is both on and off” which leads to a contradiction and thus has type $\bot$. For a physics example, consider the statement $Q$ that an object is both at rest and accelerating under a net force has type $\bot$, since it contradicts Newton's laws. This can be expressed as $\Gamma_{\text{Newton}} \vdash Q : \bot$.

\subsection{Equation rewrite form}

The symbol $\vert$ indicates the equation-rewrite form and means that, in $t \ \vert \ r$, the term on the left is rewritten using the equations on the right. Here, $t \ \vert \ r$ reads as ``$t$ under $r$”. The present paper adopts this metallanguage notation\footnote{“Metalevel” refers to the explanatory metalanguage used to talk about the formal system. The symbol $|$ belongs to this meta-language; it is not part of the object language (terms and types). This convention carries no additional type-theoretic content.} for its usefulness in symbolic manipulation, but is not standard in type theory.

\section{Cognitive typing rules} \label{sec:methods}

In the previous section, we presented various examples involving fully formed concepts—those already taken as given—to illustrate how type theory works. Now, we turn to formalizing specific mechanisms of concept formation and change using type theory. We refer to the typing rules that govern concept formation and change as \textit{cognitive typing rules}. From a physics perspective, these rules could also be interpreted as equations of motion for concept formation and change. In what follows, we present the cognitive typing rules needed to reconstruct the conceptual paths that led to the scientific discoveries discussed in Section~\ref{sec:results}.

\subsection{Property preservation rules}

Property preservation rules formalize reasoning patterns in which a concept produced by transforming another concept retains the same property i.e. intuitive type. The following rules apply to cases where the transformation is either a function or its inverse.

\begin{enumerate}
    \item \textbf{Forward property preservation.} This rule states that if the argument $x$ has type $A$, then the function $f(x)$ also has type $A$, as denoted below: 
    $$
    \frac{\Gamma \vdash x : A}
    {\Gamma \vdash f(x) : A}
    $$
    Intuitively, under this rule, if a concept has a certain property, applying a function $f$ to it preserves that property.

    \item \textbf{Backward property preservation.} This rule states that if the function $f(x)$ has type $A$, then its argument $x$ also has type $A$, as denoted below:
    $$    \frac{\Gamma \vdash f(x) : A}
    {\Gamma \vdash x : A}
    $$
    Intuitively, under this rule, if a concept resulting from a function has a certain property, then the concept it was applied to must also have that property.
\end{enumerate}

\subsection{Concept distinction rule}

The concept distinction rule formalizes the reasoning pattern by which identical concepts with different intuitive types are distinguished from one another: one retains its type, and the other is marked inconsistent.

\textbf{Concept distinction.} This rule states that if two terms, $x$ and $y$, are assigned different types $A$ and $B$ in separate contexts $\Gamma_1$ and $\Gamma_2$, and are assumed to be heterogeneously equal, in the combined context $\Gamma_1 \cup \Gamma_2$, $x$ retains its type and $y$ is assigned the empty type $\bot$, indicating an inconsistency.    
$$
\frac{\Gamma_1 \vdash x : A \quad \Gamma_2 \vdash y : B \quad x = y}
{\Gamma_1 \cup \Gamma_2 \vdash x : A \quad \Gamma_1 \cup \Gamma_2 \vdash y : \bot}
$$

Intuitively, this rule captures the line of reasoning that if two concepts are considered identical but have different properties, then only one of them can exist. Notably, this typing rule differs from the conventional ones found in Section~\ref{sec:background}, as it involves defining terms (concepts) in different contexts and then uniting those contexts.

For a common-sense example, consider two books with the same title placed in both the science fiction and history sections of a library.
According to the distinction rule, under the combined context of both sections, the librarian might assume the two entries refer to the same book and treat the differing classifications as inconsistent. Thus, the librarian might reasonably—but perhaps mistakenly—conclude that it belongs to science fiction and that its classification as history is an error. 

\subsection{Concept change rule}

The concept change rule captures the reasoning pattern that if two concepts have different intuitive types yet are identical, one acquires the other’s type as well.

\textbf{Concept change.} This rule states that if two terms, $x$ and $y$, assigned different types $A$ and $B$ in separate contexts $\Gamma_1$ and $\Gamma_2$, are assumed to be heterogeneously equal, then in the combined context $\Gamma_1 \cup \Gamma_2$, the term $x$ changes its type to the intersection type $A \cap B$.

$$
\frac{\Gamma_1 \vdash x : A \quad \Gamma_2 \vdash y : B \quad x = y}{\Gamma_1 \cup \Gamma_2 \vdash x : A \cap B}
$$

Intuitively, this rule captures the line of reasoning that if two concepts are considered identical but have different properties, then they can be treated as a single concept that possesses both properties. For a common-sense example of this rule, take avocado, which is considered savory in Peru and sweet in Brazil. Thus, depending on the cultural context, avocado may be assigned the type savory or sweet. Nonetheless someone who has experienced both cultures might treat avocado as both savory and sweet—say, by eating an avocado sandwich alongside an avocado smoothie. Notably, the rule implies that the name “avocado” does not determine the concept’s meaning; it merely serves as a pointer that locates the concept. Instead, the meaning of avocado is encoded in how the concept is used—specifically, in how it relates to other concepts, as defined by the types it is assigned and the rules that govern them.

\section{Conceptual paths}\label{sec:results}

In this section, we build on the type theory background developed in Section~\ref{sec:background} and the cognitive typing rules introduced in Section~\ref{sec:methods} to formalize the conceptual paths that led Einstein to the impossibility of frozen waves and the relativity of time. We formalize these two paths, respectively, as instances of two discovery mechanisms: \texttt{distinction} and \texttt{assumption}–\texttt{conclusion} \texttt{switch}. To clarify the kinds of concepts formalized along these conceptual paths, we begin by recalling that in Section~\ref{sec:background}, we applied typing rules to already formed common-sense and scientific concepts, and in Section~\ref{sec:methods}, to in-formation common-sense concepts that were not yet fully characterized. Here, we apply them to concepts that are both scientific and still in formation—meaning they are not yet physically correct (i.e., they wouldn’t appear in a textbook and may be forgotten from history)—but nonetheless serve as necessary provisional steps toward physically valid ones.

\subsection{Conceptual path to frozen light waves impossibility} 

\subsubsection{Setting the stage}

At the age of sixteen, Einstein wondered what a light wave would look like if one were to travel on top of it. On the basis of his experience\footnote{Contrary to the popular belief that Einstein derived this thought experiment from Maxwell’s equations, he himself admitted that he only learned Maxwell’s electrodynamics during his university years \citep{norton2004einstein}}—and perhaps influenced by Galilean relativity—he reasoned that if he traveled at the speed of light, the wave would appear frozen. Yet, he instinctively felt that a frozen wave could not be physically real. This led him to suspect that any theory permitting such a possibility had to be, at the very least, partially flawed.

Later, after learning Maxwell’s electrodynamics, Einstein likely used a thought experiment to rule out the possibility of frozen waves\footnote{as a first step in the process of ruling out emission theories of light}, as suggested by John Norton \citep{norton2005einsteineasy, norton2004einstein}. The core result of this thought experiment is that frozen waves cannot exist (or at least are not likely to) because, in standard Maxwell electrodynamics,  there is no way to distinguish between frozen and propagating waves at a single \textit{instant}\footnote{Einstein would have considered time at a single instant in his thought experiment, since that is how his claims about propagating and frozen waves would have been tested experimentally \citep{norton2004einstein}}. The following is a type-theoretic reconstruction of Einstein's reasoning in this thought experiment, showing how he arrived at this conclusion, along with an explanation in natural language:

\subsubsection{\texttt{Distinction} discovery mechanism}

\begin{enumerate}[label=\textbf{Step \arabic*.}, leftmargin=2.5cm]
    \item Einstein first considered a propagating wave, which we can formalize \\ as having the type $\texttt{move}$
    $$
    \hspace*{-3cm} E_{\text{prop}} = E_0 \sin(\omega t - ky) : \texttt{move}
    $$

    \item To examine the wave at an instantaneous moment, he evaluated it at \\ an arbitrary specific point in time, such as $t=0$
    $$
    \hspace*{-3cm} E_{\text{prop}} = E_0 \sin(-ky) : \texttt{move}
    $$
    
    \item Next, he considered a frozen wave, which is a wave constrained by two \\ key assumptions:
    \begin{itemize}
        \item It undergoes a Galilean transformation.
        \item The observer moves at the speed of light.
    \end{itemize}    
    Under these conditions, the wave is formalized as having the type $\texttt{frozen}$
    $$
    \hspace*{-3cm} E_{\text{frozen}} = E_0 \sin(\omega t - ky) \mid y = y' + vt, \ \ t = t', \ \ v = c \texttt{: frozen}
    $$

    \item Applying the transformation and velocity condition into the frozen wave
    $$
    \hspace*{-3cm} E_{\text{frozen}} = E_0 \sin(\omega t' - k(y' + vt')) \mid v = c \texttt{: frozen}
    $$
    $$
    \hspace*{-3cm} = E_0 \sin(\omega t' - k(y' + ct')) \texttt{: frozen}
    $$

    \item Einstein substituted the fact that $\omega = kc$ in the frozen wave
    $$
    \hspace*{-3cm} E_{\text{frozen}} = E_0 \sin(kct' - ky' + kct') \texttt{: frozen}
    $$
    $$
    \hspace*{-3cm} = E_0 \sin(- ky') \texttt{: frozen}
    $$

    \item Since propagating and frozen waves were understood by Einstein to be physically distinct, but he found them to be mathematically equivalent, he could have concluded that they are actually physically the same. This possible judgment can be formalized using the concept change rule:
    $$
    \hspace*{-3cm} \frac{\Gamma_1 \vdash E_{\text{prop}} : \texttt{move} \quad \Gamma_2 \vdash E_{\text{frozen}} : \texttt{frozen} \quad E_{\text{prop}} = E_{\text{frozen}}}{\Gamma_1 \cup \Gamma_2 \vdash E_{\text{prop}} : \texttt{move} \cap \texttt{frozen}}
    $$
    where
    $$
    \hspace*{-3cm}
    \Gamma_1 = \{ E_{\text{prop}} = E_0 \sin(\omega t - ky) : \texttt{move}, \quad \omega = kc \}$$
    $$
    \hspace*{-3cm}
    \Gamma_2 = \{  E_{\text{frozen}} = E_0 \sin(\omega t - ky) \mid y = y' + vt, \ \ t = t', \ \ v = c \texttt{: frozen}, \quad \omega = kc \}$$

    Nonetheless, he did not take this approach; instead, he remained steadfast in the conviction he had developed at the age of 16
    
    \item Since he knew that propagating waves exist, he would have ruled out the possibility of frozen waves to reconcile the inconsistency between the mathematics described above and his experience. This judgment can be formalized using the distinction rule:
    $$
    \hspace*{-3cm} \frac{\Gamma_1 \vdash E_{\text{prop}} : \texttt{move} \quad \Gamma_2 \vdash E_{\text{frozen}} : \texttt{frozen} \quad E_{\text{prop}} = E_{\text{frozen}}}{\Gamma_1 \cup \Gamma_2 \vdash E_{\text{prop}} : \texttt{move} \quad \Gamma_1 \cup \Gamma_2 \vdash E_{\text{frozen}} : \bot}
    $$
\end{enumerate}

The distinction discovery mechanism\footnote{This differs from the concept distinction rule in the cognitive typing rules section. Here the equality is between functions of variables (e.g., $f(x)$ and $g(y)$), not between variables themselves.} can be stated in its most general form, without referring to the example’s specific variable names, as follows:

% Define a bright thin blue
\definecolor{niceblue}{RGB}{0, 102, 204} % match from your image

\setlength\fboxsep{10pt}      % increase padding (was 2pt)
\setlength\fboxrule{0.4pt}           % thickness of the frame

% then in your document:

\begin{center}
\fcolorbox{niceblue}{white}{%
  \( \displaystyle
    \frac{\Gamma_{1}\vdash x:A \quad \Gamma_{2}\vdash y:B \quad f(x)=g(y)}
         {\Gamma_{1}\cup\Gamma_{2}\vdash x:A \quad \Gamma_{1}\cup\Gamma_{2}\vdash y:\bot}
  \)
}
\end{center}

where $f$ and $g$ are the functions that substitute the appropriate context-dependent algebraic constraints into $x$ and $y$, respectively.

\subsection{Conceptual path to the relativity of time}

The Michelson–Morley experiment is often regarded as one of the crucial optical experiments that Einstein sought to explain in his effort to extend the principle of relativity to electrodynamics. However, according to Shankland’s account of a conversation he had with Einstein, the physicist noted that he only became aware of the Michelson–Morley experiment after publishing his paper on special relativity, and that it was actually two other experiments that played this formative role:

\begin{displayquote}
“Otherwise,” he said, “I would have mentioned it in my paper.” He continued to say the
experimental results which had influenced him most were the observations of stellar
aberration and Fizeau’s measurements on the speed of light in moving water. “They were
enough,” he said.
\end{displayquote}

What role might stellar aberration and Fizeau’s experiment have played in shaping Einstein’s path to special relativity? John Norton \citep{norton2004einstein} argues that a close examination of how these results fit into Lorentz’s treatment of electrodynamics, which Einstein based his reasoning on, suggests that these results provided Einstein with empirical support for believing in the relativity of time. Since Einstein analyzed both experiments similarly, we will focus solely on explaining how he reasoned from stellar aberration to the relativity of time, and ultimately on formalizing that reasoning process using typing rules.

\begin{figure}[htbp]
    \centering
    \includegraphics [width=0.9\textwidth]{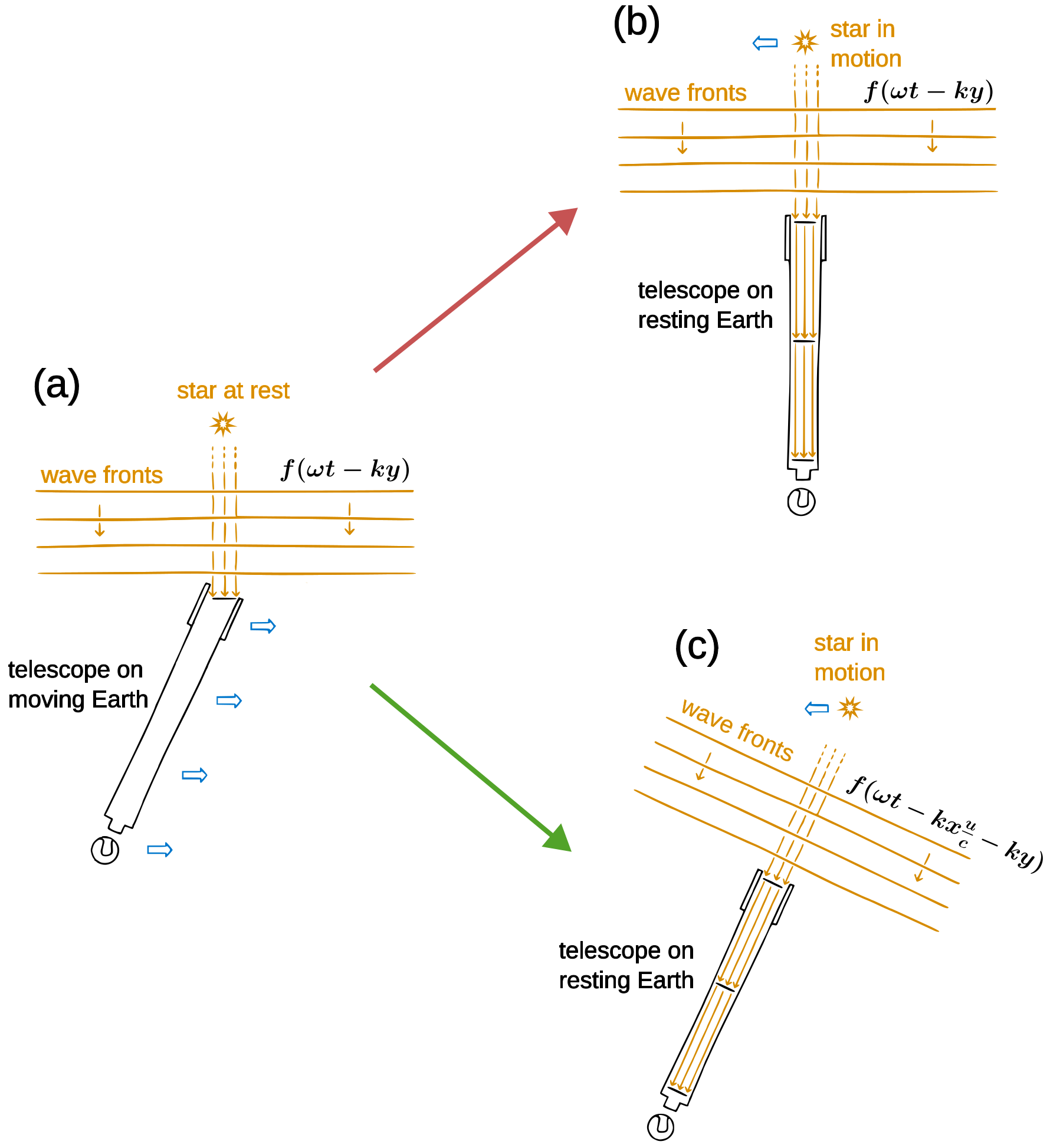} % Adjust width as needed
    \caption{Stellar aberration in the star’s rest frame and in the Earth’s rest frame under Galilean and Lorentz transformations. (a) In the star’s rest frame, Earth moves right and the telescope must tilt in order to capture light wave fronts due to stellar aberration. (b) In the Earth’s rest frame obtained via a Galilean transformation, the wavefronts remain vertical and no tilt is needed—implying the absence of stellar aberration. (c) In the Earth’s rest frame obtained via a Lorentz transformation, the wavefronts tilt, restoring the need for telescope tilt and thereby preserving stellar aberration. The images used here are adapted from John Norton’s website, with permission (personal communication).}
    \label{fig:stellar_aberr_galilean_lorenz_transfs}
\end{figure}

\subsubsection{Setting the stage}

First observed by James Bradley in 1727, stellar aberration is the apparent deflection in the direction of a star's light due to the Earth’s motion around the Sun\footnote{To help visualize this, a useful analogy of this effect is how raindrops appear to fall at a tilted angle when observed by someone walking.}. Einstein might have recognized in stellar aberration a well-established experiment involving both light and relative motion, and therefore a natural setting in which to pursue his extension of the principle of relativity to electrodynamics \citep{norton2004einstein}. For his first attempt, Einstein most likely relied on the Galilean transformation, but ultimately concluded that it could not succeed. This incompatibility becomes evident when one examines how a light wave behaves under this transformation, as we do now, following Norton’s analysis \citep{norton2005einsteineasy, norton2004einstein}.

We begin by examining light propagation in the star’s rest frame shown in Fig.~\ref{fig:stellar_aberr_galilean_lorenz_transfs}(a). The star emits light waves toward the Earth along the $y$-axis, which are denoted as $f(\omega t - ky)$. Since light wave fronts reaching Earth are virtually flat, they are plane and perpendicular to their direction of propagation. These wave fronts then reach a telescope on Earth, which moves to the right along with the Earth's motion. To ensure the starlight passes through the telescope and reaches the observer’s eyepiece, the telescope must be tilted. If left upright, the trailing wall of the telescope would block the light before it reaches the eyepiece. This necessary tilt deflects the observed direction of the starlight by an amount precisely equal to stellar aberration's deflection angle $\theta = \dfrac{u}{c}$, where $u$ is the Earth's speed as it moves around the sun and $c$ is the speed of light.

% one might wonder how an
% ether-based, wave theory of light could possibly accommodate Bradley’s result. Yet it turns out to be quite easy, as is shown in Figure 13. 

For one seeking to ensure that stellar aberration is consistent with Galilean relativity, the effect should also be observed in the Earth's rest frame shown in Fig.~\ref{fig:stellar_aberr_galilean_lorenz_transfs}(b). However, this is not the case. By applying a Galilean transformation to the star’s rest frame, we arrive at the Earth's rest frame and assign the star and its emitted light a velocity of $-u$ along the $x$-axis. Under this transformation, the coordinates change as $t' = t$, $x' = x + ut$, and $y' = y$. Consequently, the transformed wave remains $f(\omega t' - k y')$, indicating that its propagation direction is unchanged. As a result, the wave fronts remain perpendicular to the line connecting the star and Earth rather than tilting. This implies that if the star, rather than the Earth, is in motion, there is no need to tilt the telescope on Earth to observe the starlight. Thus, the effect of stellar aberration disappears when the star moves instead of the Earth, contradicting relativity.

To resolve this contradiction, Einstein sought an explanation of stellar aberration that depended only on the relative velocity between Earth and star, rather than on their individual velocities \citep{norton2004einstein}. He recalled from his reading of \textit{Versuch}, how Lorentz had applied his transformations for fields\footnote{The transformations of the field quantities turn out to play no role in the final result. All that matters to track the velocity of a wave are the locations at which the field intensity drops to zero; these are unaffected by
the field transformations. As a result, I can still simply represent a propagating wave as $f(\omega t - ky)$ where $f$ stands for the multivalued field intensities.} and coordinates to stellar aberration, as illustrated in Fig~\ref{fig:stellar_aberr_galilean_lorenz_transfs}(c). Similar to the Galilean transformation, the Lorentz transformation, when applied to the star’s rest frame, brings Earth to rest while setting the star in motion in the opposite direction. However, under the Lorentz transformation, the coordinates change as follows: $t' = t-\frac{ux}{c^2}$ and $x'= x-ut$ where $t'$ is what Lorentz referred to as the \textit{local time}—an artificial but necessary coordinate that Lorentz did not consider to represent real time. As a result, the waveform transforms into $f(\omega t'-ky')
= f(\omega (t-\frac{ux}{c^2}) - ky) = f(\omega t - k \frac{ux}{c} - ky)$, where $c=\omega/k$ and $u$ is the Earth and star's relative velocity. Notably, this transformation deflects the wave fronts' propagation direction by an angle of $\frac{u}{c}$ radians, aligning it with the star's motion. That is, the Lorentz transformation rotates the wavefronts, making them no longer perpendicular to the line connecting the star and the Earth. This means that the telescope on Earth must be tilted for starlight to reach the eyepiece. Thus, stellar aberration is preserved in the Earth's frame as it depends solely on the relative motion between the Earth and the star—a result fully consistent with the principle of relativity.

Here, Norton \citep{norton2004einstein} proposes that Einstein, by viewing Lorentz’s application of his transformations to stellar aberration through the lens of relativity, would have followed a line of reasoning leading to the conclusion that stellar aberration provided experimental support for the physical reality of the local time—allowing it to be interpreted simply as time itself. This thought process can be regarded as a discovery mechanism in which the roles of assumption and conclusion are switched. It is this mode of discovery that we describe in natural language, alongside its corresponding type-theoretic representation.

\subsubsection{\texttt{Assumption-conclusion switch} discovery mechanism}

\begin{enumerate}[label=\textbf{Step \arabic*.}, leftmargin=2.5cm]
    \item Lorentz regarded the concept of local time, $t^{\prime}=t-\frac{ux}{c^2}$, as a mathematical coordinate without physical interpretation, yet one that was useful for explaining stellar aberration while treating light as an electromagnetic wave. In other words, he considered local time to be a necessary artificial construct (\citep{lorentz1895versuch}, p. 77). This seemingly qualitative remark can be formalized as the assertion that $t^{\prime}$ has the type \texttt{art} (artificial construct), denoted as 
    $$
    \hspace*{-3cm}
    t^{\prime} \texttt{ : art}
    $$

    \item Now, a Lorentz transformation applied to stellar aberration yields a light waveform in the Earth’s rest frame, in which time is replaced by local time as follows:
    $$
    \hspace*{-3cm}
    f\left( \omega t - ky \right) \rightarrow f\left( \omega \left(t - \frac{u x}{c^2}\right) - ky \right) 
    $$   
    Here, Einstein might have wondered: what does the artificial nature of local time imply for the transformed wave? Or in our language, what does the local time’s type imply about the type of the transformed wave? The simplest answer—and the one Einstein seems to have adopted—is that anything derived from an artificial construct is itself artificial. This reasoning can be formalized by applying the forward property preservation rule to $t'$, as shown below:
    $$
    \hspace*{-3cm}
    \frac{\Gamma \vdash t^{\prime}\texttt{ : art}}{\Gamma \vdash g(t^{\prime})\texttt{ : art}}
    $$    

    Thus,
    $$
    \hspace*{-3cm}
    f\left( \omega \left(t - \frac{u x}{c^2}\right) - ky \right) \texttt{: art}
    $$
    
    \item Simplifying expression and using $k = \frac{w}{c}$, we have
    $$
    \hspace*{-3cm}
    = f\left( \omega t - \omega \frac{u x}{c^2} - ky \right) \texttt{: art}
    $$
    $$
    \hspace*{-3cm}
    = f\left( \omega t - k x \frac{u}{c} - ky \right) \texttt{: art}
    $$
    \item Note that we can summarize the previous steps as follows: Under the assumption that local time is an artificial construct, the transformed waveform must also be artificial. This can be formalized by saying that under the context $\Gamma_1 = t^{\prime}$\texttt{ : art}, we have
    $$
    \hspace*{-3cm}
    \Gamma_1 \vdash f\left( \omega t - k x \frac{u}{c} - ky \right) \texttt{: art}
    $$
    To simplify notation, we define the term above as $f_{\text{art}}$. Thus,
    $$
    \hspace*{-3cm}
    \Gamma_1 \vdash f_{\text{art}} \texttt{: art}
    $$
    \item On the other hand, Einstein recognized that, in the Earth’s rest frame, the light waveform deflected by stellar aberration $f( \omega t - kx\frac{u}{c} - ky)$ is a phenomenological observation that does not rely on theoretical assumptions. Thus, the waveform had to be empirically true rather than artificially constructed. To formalize this reasoning, we state that, under the empty context $\Gamma_2 = \varnothing$, the deflected waveform has type $\texttt{emp}$ (empirical):
    $$
    \hspace*{-3cm}
    \Gamma_2 \vdash f\left( \omega t - kx\frac{u}{c} - ky \right) \texttt{: emp}
    $$
    For convenience, we define the term above as $f_{\text{emp}}$. Thus,
    $$
    \hspace*{-3cm}
    \Gamma_2 \vdash f_{\text{emp}} \texttt{: emp}
    $$
    \item Here, Einstein would have recognized that although $f_{\text{art}}$ is artificial, it is also equal to $f_{\text{emp}}$, which is empirical. According to Norton's analysis, this suggests that, at the very least, $f_{\text{art}}$ could not be solely artificial but must also possess an empirical aspect. Thus, within our formalism, Einstein would have concluded that $f_{\text{art}}$ has both types: \texttt{art} and \texttt{emp}. This reasoning process can be formalized using the concept change rule:
    $$
    \hspace*{-3cm}
    \frac{\Gamma_1 \vdash f_{\text{art}} \texttt{ : art}  \quad \Gamma_2 \vdash f_{\text{emp}}\texttt{: emp} \quad f_{\text{art}} = f_{\text{emp}}}{\Gamma_1 \cup \Gamma_2 \vdash f_{\text{art}}\texttt{ : art} \ \cap \ \texttt{emp}}
    $$
    where
    $$
    \hspace*{-3cm}
    \Gamma_1 = \{ t^{\prime}\texttt{ : art} \} \quad
    \text{and} \quad 
    % \hspace*{-3cm}
    \Gamma_2 = \varnothing
    $$
    \item At this point, according to Norton \citep{norton2004einstein}, Einstein reasons that since his earlier conclusion is not merely artificial but also empirically true, he can now reinterpret it as an assumption. In terms of our formalism, Einstein assumes that $f_{\text{art}}$ is of both types: $\texttt{art}$ and $\texttt{emp}$. As a result, he begins to derive the previous steps in reverse order, while cautiously maintaining that the same type ($\texttt{art} \ \cap \ \texttt{emp}$) continues to propagate backwards. 
    
    Eventually, by reasoning backwards, Einstein reaches a new version of his former assumption, now framed as a conclusion: local time $t'$ has type $\texttt{art} \ \cap \ \texttt{emp}$. That is, local time could not be wholly artificial; it had to be, at least in part, empirically true. In other words, local time could be time itself, not merely an artifice. Einstein's reasoning process can be formalized using a backward property preservation rule as follows:
    $$
    \hspace*{-3cm}
    \frac{\Gamma_1 \cup \Gamma_2 \vdash f_{\text{art}}\texttt{ : art} \ \cap \ \texttt{emp}}{\Gamma_1 \cup \Gamma_2 \vdash t^{\prime}\texttt{ : art} \ \cap \ \texttt{emp}}
    $$
    
    Because local time depends on relative velocity, it was at this moment that Einstein realized time could be relative. That realization launched special relativity.    
    \item Ultimately, Einstein's full discovery mechanism of switching Lorentz' assumption and conclusion can be formalized using the following rule:
    $$
    \hspace*{-3cm}
    \frac{\Gamma_1 \vdash t^{\prime}\texttt{ : art}  \quad \Gamma_2 \vdash f_{\text{emp}}\texttt{: emp} \quad g(t^{\prime}) = f_{\text{emp}}}{\Gamma_1 \cup \Gamma_2 \vdash t^{\prime}\texttt{ : art} \ \cap \ \texttt{emp}}
    $$
    
    That is, under assumptions from contexts $\Gamma_1$ and $\Gamma_2$, the concept of local time $t^{\prime}$ is assigned both types: \texttt{art} and \texttt{emp}. This rule builds compositionally upon the concept change and property preservation rules along with other simple mathematical operations.

\end{enumerate}

The assumption-conclusion switch discovery mechanism can be stated in its most general form, without referring to the example’s specific variable names, as follows:

% Define a bright thin blue
\definecolor{niceblue}{RGB}{0, 102, 204} % match from your image

\setlength\fboxrule{0.4pt}           % thickness of the frame
\setlength\fboxsep{10pt}      % increase padding (was 2pt)

% then in your document:
\begin{center}
\fcolorbox{niceblue}{white}{
  \( \displaystyle
    \frac{\Gamma_{1}\vdash x : A \quad \Gamma_{2}\vdash y : B \quad g(x)=y}
         {\Gamma_{1}\cup\Gamma_{2}\vdash x : A \cap B}
  \)
}
\end{center}

where $g$ denotes the function that transforms $x$ into $y$.

Note that one might expect this chain of reasoning to conclude that Einstein became fully aware that time is relative, that is, that $t'$ is solely of type $\texttt{emp}$ ($t' \ \texttt{: emp}$). In fact, this conclusion could be reached by applying a version of the distinction distinction discovery mechanism. However, making that claim would be premature. By this point Einstein had only just begun to regard the relativity of time as a reasonable possibility and was not yet fully confident. Further work could apply Bayesian reasoning to assign probabilities over the terms in the concept-change rule, with outcomes that depend on a ``boldness'' prior.

\subsection{Bridge from Type theory to Computation}

To carry the type-theoretic formalism into a computational setting, we examine the three-way relationship among type theory, programming languages and scientific concepts introduced in Section~\ref{sec:intro}. We begin with the connection between type theory and programming languages. Specifically, we note that once a type system is equipped with fixed operational semantics (step-by-step evaluation rules), the system's terms can be viewed as programs and its theoretical types as programming-language types \citep{Plotkin1977LCF}.

Turning to the link between programming languages and scientific concepts, recent work in computational cognitive science holds that symbolic programs (code) offer the best formal representation of concepts because of their expressive power \citep{rule2020child}. As noted in Section~\ref{sec:intro}, Sosa builds on this perspective by suggesting that types in programming languages may capture the cognitive constraints that impose structure in concepts and allow us to generate reasonable, though sometimes incorrect, answers \citep{sosa2022type}. Motivated by these three developments, we introduce two analogies—one theoretical and one computational—for formalizing scientific concept formation.

\definecolor{cellbg}{RGB}{230,245,255}
\definecolor{darkblue}{RGB}{0,0,139}
% \definecolor{ruleblue}{RGB}{0,0,150}
\definecolor{niceblue}{RGB}{0, 102, 204} % match from your image
\definecolor{nicedarkblue}{RGB}{0,  82, 164}  % about 20% darker

\begin{tcolorbox}[
  colback=white,
  colframe=niceblue,
  boxrule=0.6pt,
  arc=0pt,
  left=6pt,right=6pt,top=6pt,bottom=6pt
]
\begin{enumerate}
  \item \textbf{Theoretical analogy:} Intuitive types are formalized as theoretical types; discovery mechanisms as typing rules; and scientific concepts as terms and the conclusions of typing rules.
  \item \textbf{Computational analogy:} Intuitive types are formalized as types in a programming language; discovery mechanisms and scientific concepts as programs.
\end{enumerate}
\end{tcolorbox}

These analogies provide the basis for the Python implementation of the type-theoretic reconstruction of Einstein’s conceptual path to the relativity of time.

\section{Program synthesis}
\label{sec:program-induction}

In the previous sections, we introduced a type-theoretic reconstruction of scientific concept formation. In this section, the reconstruction is recast as a program-synthesis task: the inference of programs given constraints. Throughout, Einstein’s conceptual path to the relativity of time serves as the running example.

To carry out this recasting, we implement the type-theoretic formalism (terms, types, and typing rules) as executable code. To ensure the implementation is user-friendly and practically scalable, we choose to implement the formalism in Python rather than any domain-specific language (DSL), including DSLs that become Turing complete when equipped with an escape hatch \citep{EllisTavaresProgSynthVideo}. In what follows, each formal element is mapped to a Python counterpart.

\subsection{Representing Type-theoretic objects in Python}

Terms are represented as Python strings so that they can be algebraically manipulated with ease. As strings, each term can be passed directly to SymPy—Python’s symbolic-mathematics library—which then parses it into a symbolic object that can undergo substitution, simplification, and other operations. This representational choice naturally supports the algebraic manipulations involved in Einstein’s conceptual path to the relativity of time. Furthermore, to keep manipulations concise, we sometimes assign a name to a term. Each name is definitionally equivalent to the term it stands for.

Intuitive types are represented as instances of Python classes. This mirrors how Python handles both built-in and user-defined types—as instances of the built-in class \texttt{Type} or an explicitly defined Python class. Specifically, Einstein's intuitive types— \texttt{Art}, \texttt{Emp}, and \texttt{Intersection}—are implemented as instances of the class \texttt{Base}. Similarly, contexts are implemented as instances of the class \texttt{Context}, and each such instance is represented as a set of strings.

A judgment is represented as an instance of the class \texttt{Judgment}, consisting of four elements: context, name, term and type. Moreover, typing rules are implemented as Python functions that take an instance of the \texttt{Judgment} class as input and return a new instance of the same class as output. Notably, a backward property preservation rule simply goes back to whathever term we had immediately before; however, what we call forward property preservation rule is constrained to be a function application $f$, instead of any other algebraic operation or cognitive typing rule.

\subsection{State, Goal and Solution}

To make the implementation explicit, we frame scientific concept formation as the process of transforming an initial state into a final state that satisfies a specified goal. Within this framing, a state comprises either a single judgment or a set of judgments. The initial state is a set of judgments, and the final state is a single judgment. For simplicity, the goal is defined to be the final state itself. Hence, a solution is a sequence of operations and typing rules that transform the initial state into the final state.

The initial state consists of two judgments: the local time $t - \frac{u x}{c^{2}}$ has type \texttt{art}, and the light waveform exhibiting stellar aberration has type \texttt{emp}. The final state, which also serves as the goal, is the judgment that local time $t - \frac{u x}{c^{2}}$ is of type \texttt{emp}. The goal of our present implementation is simple: recover the ground-truth judgment supplied by an oracle. In future implementations, a more cognitively and historically realistic goal would be to find judgments that satisfy a given constraint on the final state, or to construct new relationships among the existing judgments. With the states, goal, and solution now clearly specified, we can now introduce the search procedures.

\subsection{Search}

\begin{figure}[htbp]
    \centering
    \includegraphics [width=1.0\textwidth]{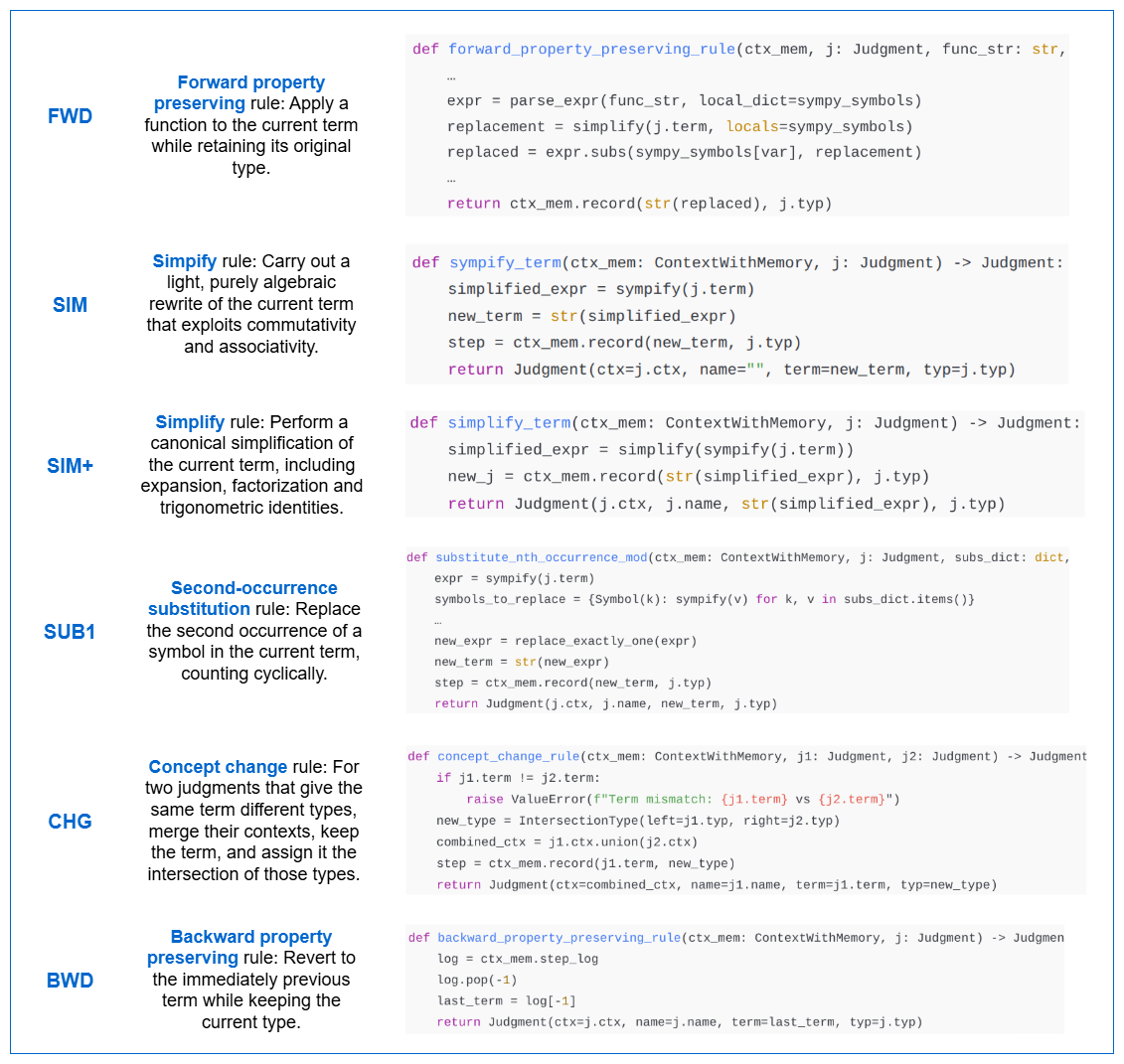} % Adjust width as needed
    \caption{Primitives that serve as the building blocks of the search. The left column names each primitive, the middle explains what each one does in natural language, and the right column shows a shortened version of the Python implementation of each (for illustration).}
    \label{fig:einstein_relative_time_primitives}
\end{figure}

For researchers working at the intersection of AI and the natural sciences, an intuitive way to understand the search conducted by our program-synthesis approach is to view it by analogy to the more familiar symbolic regression. Both approaches explore expressions built from primitives. In symbolic regression, those primitives are low-level arithmetic and control operators and the expressions take the form of equations, whereas in our method the primitives are typing rules and the expressions represent scientific concepts, which may, but need not, be equations. Our search relies on the primitives (mathematical operations and typing rules)\footnote{Inspired by Peano \citep{poesia2023peano}, we keep the number of judgments finite at each step by only allowing in-line function applications (not in-line function abstractions) and applying them solely to existing arguments.} explained in natural language and implemented as Python programs in Fig.~\ref{fig:einstein_relative_time_primitives}. Notably, we exclude basic mathematical operations like addition and multiplication from the primitive set, opting instead for higher-level primitives—such as algebraic simplification—that better reflect the conceptual level at which a trained physicist like Einstein reasoned, while also avoiding the combinatorial explosion caused by ultra-fine primitives.

To assess whether Bayesian and neurosymbolic guidance further mitigate the \textit{curse of compositionality} (combinatorial blow-up from composing primitives) \citep{Spelke2022WhatBabiesKnow}, three search methods are used: Enumerative, Bayesian, and Bayes-neural. These methods explore the space of eight-step sequences of mathematical operations and cognitive typing rules and identify the sequence that yields a judgment satisfying our goal. Enumerative search samples sequences by selecting primitives at each step according to a goal measure (term similarity and type agreement) and choosing the next primitive greedily or in proportion to that score. Bayesian search builds each chain by keeping a Beta$(\alpha,\beta)$ posterior for every admissible primitive (interpreted as that primitive’s chance of moving closer to the goal). At each step, for each admissible primitive $p$ we draw 
$\theta_p \sim \text{Beta}(\alpha_p, \beta_p)$ and select the primitive with the largest draw (Thompson sampling). We then update that primitive’s posterior with the goal measure. Priors are initialized $\alpha=\beta=1$ and reset between chains unless noted.

Bayesian-neural search first trains a neural policy over primitives on traces generated by the pure Bayesian method using state features (goal-term similarity, goal-type match, normalized depth). During search, each step samples the next primitive from a depth-annealed mixture of the neural policy probability and a Thompson-sampled Bayesian probability (the mixture shifts from Bayesian‑heavy in early steps to neural‑heavy in later ones). After selecting a primitive, its Beta parameters are updated with the same goal reward used in Bayesian search.

Fig.~\ref{fig:relative_time_python_flow_chart} shows the successful eight-step sequence alongside the alternative candidates for the transition from step 1 to step 2. At each step, we select one primitive from our finite pool of valid options, but not all primitives are permitted at every stage (for example, concept change is allowed only when two judgments share the same term but have different types). Moreover, for the three search methods, we can enable or disable specific heuristics to imbue the search with human-like inductive biases or maintain a purely unbiased exploration.

\usetikzlibrary{positioning,arrows.meta,bending}
\usetikzlibrary{positioning,arrows.meta}
\usetikzlibrary{arrows.meta,positioning}

% Macro to render a multi‑line Python Judgment call
\newcommand{\PyJ}[4]{%
  \shortstack[l]{%
    \texttt{Judgment(}#1,\\
    \hspace*{1.5em}#2,\\
    \hspace*{1.5em}#3,\\
    \hspace*{1.5em}#4\texttt{)}%
  }%
}

% TikZ styles
\tikzset{
  pyjudg/.style={
    rectangle,draw,rounded corners,
    inner sep=2pt,outer sep=1pt,
    font=\ttfamily\scriptsize,
    text width=46mm,
    align=left
  },
  chosen/.style={->,very thick},
  alt/.style={->,thin,dashed},
  lab/.style={font=\scriptsize,inner sep=1pt}
}

\definecolor{cellbg}{RGB}{230,245,255}
\definecolor{pykw}{RGB}{0,100,0}
\definecolor{darkblue}{RGB}{0,0,139}
\definecolor{darkpurple}{RGB}{75,0,130}
\definecolor{darkred}{RGB}{178,34,34}  % Firebrick

\begin{figure}[htbp]  
  \centering
  \begin{tikzpicture}[
    scale=0.9,
    transform shape,
    >=Latex,
    node distance=18mm and 12mm,
    every node/.style={font=\scriptsize},
    arrow label/.style={midway,yshift=3pt,fill=white,inner sep=1pt,font=\scriptsize},
    pyjudg/.style={
      rectangle,draw,rounded corners,fill=cellbg,
      inner sep=2pt,outer sep=1pt,
      font=\ttfamily\scriptsize,
      text width=46mm,align=left
    },
    chosen/.style={->,very thick},
    alt/.style={->,thin,dashed},
    lab/.style={font=\scriptsize,inner sep=1pt}
  ]

    % Step 0
    \node[pyjudg] (S0) {%
      \shortstack[l]{%
        {\color{pykw}\texttt{Judgment}}(\color{darkblue}\texttt{ctx\_lorentz},\\
        \hspace*{1.5em}{\color{darkred}\texttt{"tprime"}},\\
        \hspace*{1.5em}{\color{darkred}\texttt{"t - u*x/c**2"}},\\
        \hspace*{1.5em}{\color{pykw}\texttt{Art}}%
        )%
      }%
    };
    \node[lab,above=1mm of S0]{Step 0};

    % Step 1
    \node[pyjudg,right=of S0] (S1) {%
      \shortstack[l]{%
        {\color{pykw}\texttt{Judgment}}(\color{darkblue}\texttt{ctx\_lorentz},\\
        \hspace*{1.5em}{\color{darkred}\texttt{""}},\\
        \hspace*{1.5em}{\color{darkred}\texttt{"f(-k*y + w(t - u*x/c**2))"}},\\
        \hspace*{1.5em}{\color{pykw}\texttt{Art}}%
        )%
      }%
    };
    \node[lab,above=1mm of S1]{Step 1};

    % Step 2
    \node[pyjudg,right=of S1] (S2) {%
      \shortstack[l]{%
        {\color{pykw}\texttt{Judgment}}(\color{darkblue}\texttt{ctx\_lorentz},\\
        \hspace*{1.5em}{\color{darkred}\texttt{""}},\\
        \hspace*{1.5em}{\color{darkred}\texttt{"f(-k*y + t*w - u*w*x/c**2)"}},\\
        \hspace*{1.5em}{\color{pykw}\texttt{Art}}%
        )%
      }%
    };
    \node[lab,above=1mm of S2]{Step 2};

    % Step 3
    \node[pyjudg,below=of S2] (S3) {%
      \shortstack[l]{%
        {\color{pykw}\texttt{Judgment}}(\color{darkblue}\texttt{ctx\_lorentz},\\
        \hspace*{1.5em}{\color{darkred}\texttt{""}},\\
        \hspace*{1.5em}{\color{darkred}\texttt{"f(-k*y + t*w - k*u*x/c)"}},\\
        \hspace*{1.5em}{\color{pykw}\texttt{Art}}%
        )%
      }%
    };
    \node[lab,below=1mm of S3]{Step 3};

    % Step 4
    \node[pyjudg,left=of S3] (S4) {%
      \shortstack[l]{%
        {\color{pykw}\texttt{Judgment}}(\color{darkblue}\texttt{ctx\_combined},\\
        \hspace*{1.5em}{\color{darkred}\texttt{""}},\\
        \hspace*{1.5em}{\color{darkred}\texttt{"f(-k*y + t*w - k*u*x/c)"}},\\
        \hspace*{1.5em}{\color{pykw}\texttt{Art $\wedge$ Emp}}%
        )%
      }%
    };
    \node[lab,above=1mm of S4]{Step 4};

    % Step 5
    \node[pyjudg,left=of S4] (S5) {%
      \shortstack[l]{%
        {\color{pykw}\texttt{Judgment}}(\color{darkblue}\texttt{ctx\_combined},\\
        \hspace*{1.5em}{\color{darkred}\texttt{""}},\\
        \hspace*{1.5em}{\color{darkred}\texttt{"f(-k*y + t*w - k*u*x/c)"}},\\
        \hspace*{1.5em}{\color{pykw}\texttt{Art $\wedge$ Emp}}%
        )%
      }%
    };
    \node[lab,above=1mm of S5]{Step 5};

    % Step 6
    \node[pyjudg,below=of S5] (S6) {%
      \shortstack[l]{%
        {\color{pykw}\texttt{Judgment}}(\color{darkblue}\texttt{ctx\_combined},\\
        \hspace*{1.5em}{\color{darkred}\texttt{""}},\\
        \hspace*{1.5em}{\color{darkred}\texttt{"f(-k*y + t*w - u*w*x/c**2)"}},\\
        \hspace*{1.5em}{\color{pykw}\texttt{Art $\wedge$ Emp}}%
        )%
      }%
    };
    \node[lab,below=1mm of S6]{Step 6};

    % Step 7
    \node[pyjudg,right=of S6] (S7) {%
      \shortstack[l]{%
        {\color{pykw}\texttt{Judgment}}(\color{darkblue}\texttt{ctx\_combined},\\
        \hspace*{1.5em}{\color{darkred}\texttt{""}},\\
        \hspace*{1.5em}{\color{darkred}\texttt{"f(-k*y + w(t - u*x/c**2))"}},\\
        \hspace*{1.5em}{\color{pykw}\texttt{Art $\wedge$ Emp}}%
        )%
      }%
    };
    \node[lab,above=1mm of S7]{Step 7};

    % Step 8
    \node[pyjudg,right=of S7] (S8) {%
      \shortstack[l]{%
        {\color{pykw}\texttt{Judgment}}(\color{darkblue}\texttt{ctx\_combined},\\
        \hspace*{1.5em}{\color{darkred}\texttt{"tprime"}},\\
        \hspace*{1.5em}{\color{darkred}\texttt{"t - u*x/c**2"}},\\
        \hspace*{1.5em}{\color{pykw}\texttt{Art $\wedge$ Emp}}%
        )%
      }%
    };
    \node[lab,above=1mm of S8]{Step 8};

    % arrows
    \draw[alt]    ([yshift=1.5pt]S0.east) -- ++(1.0,1.7)  node[above,pos=0.3,yshift=12pt]{SIM};
    \draw[alt]    ([yshift=-1.5pt]S0.east) -- ++(1.0,-1.7) node[below,pos=0.3,yshift=-12pt]{SIM+};
    \draw[chosen] (S0.east) -- node[arrow label,above]{FWD}  (S1.west);
    \draw[chosen] (S1.east) -- node[arrow label,above]{SIM+} (S2.west);
    \draw[chosen] (S2.south) -- node[arrow label,right]{SUB1} (S3.north);
    \draw[chosen] (S3.west) -- node[arrow label,above]{CHG}  (S4.east);
    \draw[chosen] (S4.west) -- node[arrow label,above]{BWD}  (S5.east);
    \draw[chosen] (S5.south) -- node[arrow label,right]{BWD} (S6.north);
    \draw[chosen] (S6.east) -- node[arrow label,above]{BWD}  (S7.west);
    \draw[chosen] (S7.east) -- node[arrow label,above]{BWD}  (S8.west);

  \end{tikzpicture}
  \caption{Successful eight‑step sequence produced by the Bayesian–neural search algorithm. Each box displays a judgment consisting of a context, name, term and type. At each step, the algorithm selects exactly one primitive from the finite pool. Solid-arrow labels denote the primitive applied to move from one step to the next. Dashed arrows indicate alternative primitives; for clarity, they are shown only for Step $0 \rightarrow 1$. At the Step $3 \rightarrow 4$ transition—when CHG is applied—the judgment's context switches from $\texttt{ctx\_lorentz}$ to $\texttt{ctx\_combined}$, and its type from $\texttt{Art}$ to $\texttt{Art} \wedge \texttt{Emp}$. The goal judgment is reached at Step 8.} 
    \label{fig:relative_time_python_flow_chart}
  
\end{figure}
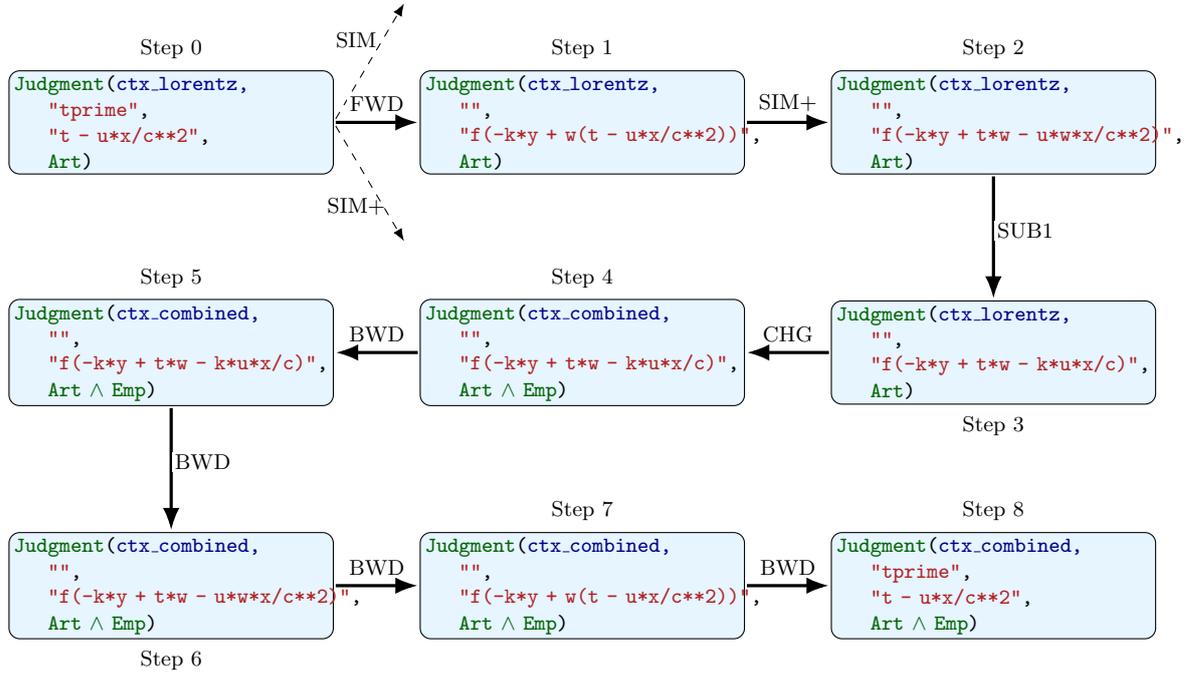

\begin{figure}[htbp]
    \centering
    \includegraphics [width=1.0\textwidth]{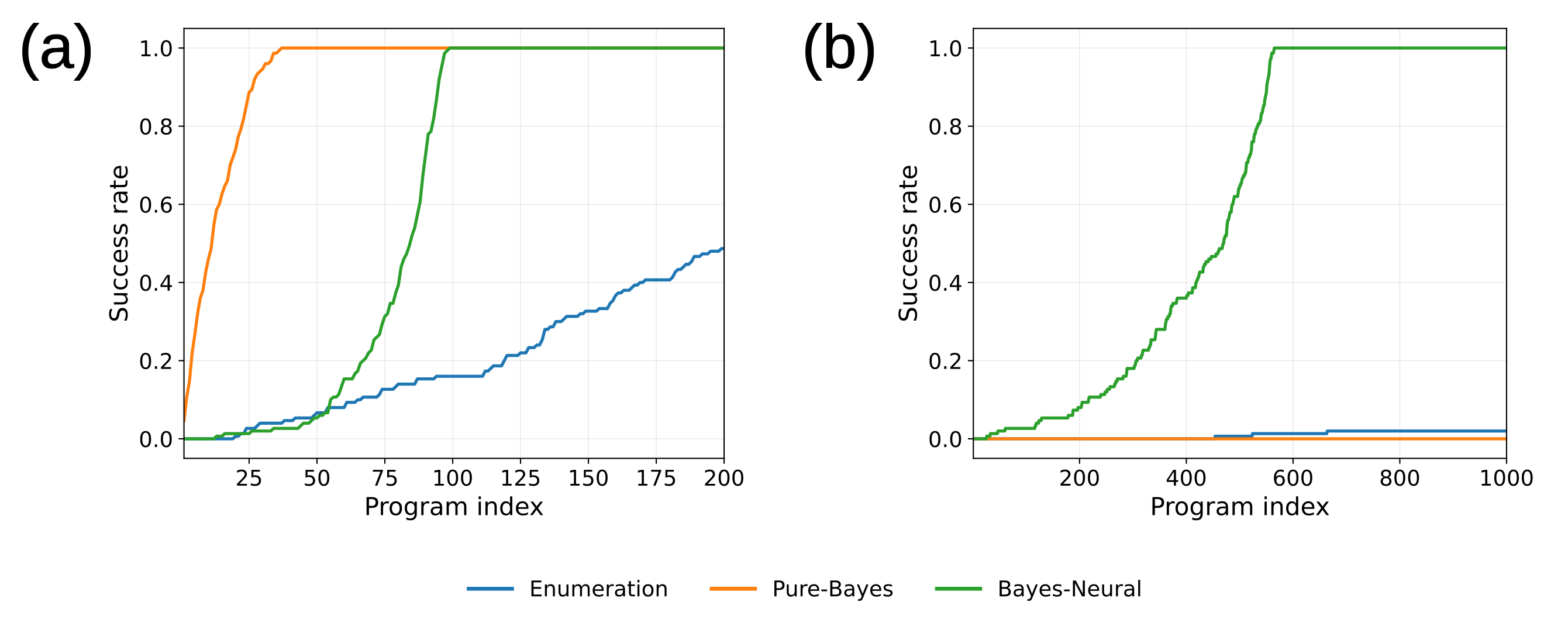}
    
    % Adjust width as needed
    \caption{Success rate (fraction of programs reaching goal so far) versus program index (number of programs evaluated) for Enumeration (blue), Pure‑Bayes (orange), and Bayes‑Neural (green) search algorithms. Panels (a) and (b) show the biased and unbiased cases, respectively. In (a) all inductive‑bias heuristics are enabled except “backward‑only after target waveform is reached”. In (b) all heuristics are off. Inductive biases greatly accelerate all search algorithms. Overall, Bayes‑Neural is the most sample‑efficient, dominating in the unbiased case and remaining near the top when biases are enabled.}
    \label{fig:success_rate_over_program_indices}
\end{figure}

Fig.~\ref{fig:success_rate_over_program_indices}  plots the success rate of each of the three search methods versus program index (i.e., the number of programs run up to that point), averaged over $N=150$ independent runs. For each method, the text reports when $100\%$ success is first reached as “about $X$ programs (median: $M$, IQR: $I$)." Fig.~\ref{fig:success_rate_over_program_indices}(a) shows results with all search constraints enabled except the  “backward‑only after target waveform is reached” inductive bias. Under these constraints, Pure‑Bayes achieves the highest performance, Bayes‑Neural the next best, and Enumeration the lowest. Pure‑Bayes reaches $100\%$ success at about $35$ programs (median: 13, IQR: 17), Bayes‑Neural does so by about $98$ programs (median: 88.5, IQR: 22.75), and Enumeration never attains full success within the $200$‑program budget. When all search constraints are removed, Fig.~\ref{fig:success_rate_over_program_indices}(b) shows Bayes‑Neural dramatically surpasses Pure‑Bayes and Enumeration. Bayes-Neural converges to $100\%$ success at about $558$ programs (median: 485, IQR: 107.5), whereas Pure‑Bayes and Enumeration remain below $3\%$ success even after $1000$ programs. The pseudocode for the Bayesian-neural search, the strongest performer overall, is shown in Algorithm~\ref{alg:bayes_neural_algorithm}.

\begin{algorithm}[t]
\caption{Bayes--Neural Hybrid Search (Python-style pseudocode)}
\label{alg:bayes_neural_algorithm}
\begin{algorithmic}[1]

\Procedure{RunExperiments}{$initState,\,policy,\,numRuns$}
  \State $results \gets [ \ ]$
  \For{$i=1$ \textbf{to} $numRuns$}
    \State $results.\mathrm{append}\bigl(\Call{BayesNeuralSearch}{initState,\,policy}\bigr)$
  \EndFor
  \State \Return $results$
\EndProcedure

\Procedure{BayesNeuralSearch}{$initState,\,policy$}
  \State $state \gets initState$
  \ForAll{$p \in PRIMITIVES$} \State $\alpha_p,\beta_p \gets 1,1$ \EndFor
  \State $flags \gets \varnothing$
  \For{$step=1$ \textbf{to} $L$}
    \State $\mu \gets MIX\_BASE + (1-MIX\_BASE)\cdot(step-1)/(L-1)$ \Comment{depth anneal}
    \State $\pi \gets policy(\Call{EncodeState}{state})$ \Comment{neural policy probs}
    \State $candidates,weights \gets [ \ ],[ \ ]$
    \ForAll{$p \in PRIMITIVES$}
      \If{$\textbf{not} \Call{ViolatesConstraints}{p,\,state,\,flags}$}
        \State $b \sim \mathrm{Beta}(\alpha_p,\beta_p)$ \Comment{Thompson draw}
        \State $candidates.\mathrm{append}(p)$
        \State $weights.\mathrm{append}(\mu\cdot\pi[p] + (1-\mu)\cdot b)$
      \EndIf
    \EndFor
    \If{$candidates=\varnothing$} \State \textbf{break} \EndIf
    \State $weights \gets \Call{Normalize}{weights}$
    \State $choice \gets \Call{Sample}{candidates,\,weights}$ \Comment{sample next primitive}
    \State $(ctx,cursor) \gets \Call{ApplyPrimitive}{choice,\,state}$
    \State $flags \gets \Call{UpdateFlags}{flags,\,choice}$
    \State $state \gets \Call{State}{ctx,\,cursor,\,state.trace \,\Vert\, choice}$
    \State $r \gets \Call{StepReward}{cursor}$ \Comment{goal closeness $r\in[0,1]$}
    \State $\alpha_{choice} \gets \alpha_{choice}+r$ \Comment{update posterior}
    \State $\beta_{choice} \gets \beta_{choice}+(1-r)$
    \If{\Call{IsGoal}{$state$}} \State \Return $step$ \EndIf
  \EndFor
  \State \Return $\textsc{None}$
\EndProcedure

\end{algorithmic}
\end{algorithm}

\subsection{Toward theory-building AI-assistants}

Viewed more broadly, typing rules and inductive biases provided us with a way to explicitly encode aspects of Einstein's discovery style and taste preferences. Looking ahead, these rules and biases could be used to encode preferences for concrete instances of other discovery strategies: favoring simple or complex analogies \citep{maxwell1865lines, HofstadterSander2013}; importing new mathematics into physics \citep{arkani-hamed2013amplituhedron, Crutchfield1989}; crafting thought experiments \citep{norton2004einstein}; or pushing a well-established theory to its ultimate consequences, à la John Wheeler’s \textit{radical conservatism} \citep{thorne2019wheeler}. In turn, this could allow us to build theory-building AI assistants that help us transfer these discovery strategies to new domains and, when appropriate, refashion them into new mechanisms, fostering conceptual leaps.

These theory-building AI assistants would serve as analogs of computer algebra systems (\citep{ellis2020algorithms}, p. 185), but for conceptual work. They would support the conceptual modeling of not well understood physical phenomena, rather than merely performing symbolic manipulations of the equations that describe them. Their goal would be to aid scientists in concept discovery, support informed conceptual risk-taking, and work alongside scientists rather than act as answering machines intended to replace them (even if not very good replacements).

% \section{Figures}

% As per the \LaTeX\ standards you need to use eps images for \LaTeX\ compilation and \verb+pdf/jpg/png+ images for \verb+PDFLaTeX+ compilation. This is one of the major difference between \LaTeX\ and \verb+PDFLaTeX+. Each image should be from a single input .eps/vector image file. Avoid using subfigures. The command for inserting images for \LaTeX\ and \verb+PDFLaTeX+ can be generalized. The package used to insert images in \verb+LaTeX/PDFLaTeX+ is the graphicx package. Figures can be inserted via the normal figure environment as shown in the below example:

%%=============================================%%
%% For presentation purpose, we have included  %%
%% \bigskip command. Please ignore this.       %%
%%=============================================%%
% \bigskip
% \begin{verbatim}
% \begin{figure}[<placement-specifier>]
% \centering
% \includegraphics{<eps-file>}
% \caption{<figure-caption>}\label{<figure-label>}
% \end{figure}
% \end{verbatim}
% \bigskip
%%=============================================%%
%% For presentation purpose, we have included  %%
%% \bigskip command. Please ignore this.       %%
%%=============================================%%

% \begin{figure}[h]
% \centering
% \includegraphics[width=0.9\textwidth]{fig.eps}
% \caption{This is a widefig. This is an example of long caption this is an example of long caption  this is an example of long caption this is an example of long caption}\label{fig1}
% \end{figure}

\section{Conclusion} \label{sec:conclusion}

So how are scientific concepts birthed? This work does not offer a final answer, but it argues that a central part of the answer lies in the intuitive types that constrain and guide scientific knowledge-making practices. By reconstructing physicists’ reasoning leading to concept discovery from philosophical analysis of historical records and formalizing this reasoning within a type-theoretic formalism, we have shown that these intuitive types, and their role in theoretical physics reasoning, can be articulated in a way that is explicit, structured, and computable. More specifically, using types and cognitive typing rules, we formalized two key episodes in Einstein’s reasoning toward special relativity: the conceptual paths that led to his rejection of frozen waves and his reinterpretation of time as relative. In doing so, we identified and formalized two discovery mechanisms he employed: \texttt{distinction} and \texttt{assumption}-\texttt{conclusion switch}. Finally, we observed that combining Bayesian methods with neural network techniques to guide search through program space effectively overcomes the curse of compositionality.

Although much remains to be understood, this framework supports the view that scientific concept discovery is not fully mystical or unformalizable, but is at least partially a human-explainable, rule-based process. The viability of this framework hinges on the idea that the vagueness often associated with defining scientific concepts can be limited by defining them in terms of how they are used in the discovery process, and by making explicit the epistemic context in which they are discovered. Furthermore, the framework suggests that history of science and philosophical insight play an essential role—one that goes beyond motivation or decoration—in formalizing scientific concept formation by shaping how formal systems and computational models are designed to represent that very process. 

More broadly, this type-theoretic approach offers a step toward a \textit{fine-grained} computational cognitive science of science—one capable of explaining the detailed reasoning involved in scientific concept discovery. Furthermore, this work complements coarse-grained cognitive formalisms that study discovery in general or at the level of research communities—stochastic or agent-based accounts and Bayesian treatments of explanatory values—rather than individual scientists’ reasoning \citep{wolpert2024stochastic, Dubova2022Against, dedeo2020bayes}. A natural next step is to investigate whether macro-level accounts of discovery strategies are recoverable from fine-grained reconstructions of individual discoveries.

Looking forward, this framework can be extended by reconstructing the discovery process in other historical examples from physics and science more broadly, formalizing them to uncover new typing rules—or to show that rules like concept distinction and concept change are reusable. In parallel, by examining research records and interviewing scientists, we may be able to formalize aspects of the scientific concept discovery process as it occurs in the present day. Moreover, the program synthesis search can be made more human-like by using more flexible constraints for the goal (e.g., satisfying specific properties rather than matching an exact equation), and by developing heuristics that are learnable and reusable rather than manually engineered \citep{poesia2023peano}. This formalization of scientific concept formation would provide us with a deeper explanation of scientific concept discovery as a phenomenon in its own right. This deeper explanation, in turn, may help us make more informed decisions about which discovery mechanisms to use in specific contexts, and avoid dismissing scientific goals grounded in mechanisms that have led to successful discoveries in the past—mechanisms we might have forgotten. Thus, a type-theoretic formalism of scientific concept formation may serve as both tool AI \citep{aguirre2025engineering} and a supportive framework for scientists—especially those early in their careers—to take \textit{conceptual risks} they might otherwise hesitate to pursue. Ultimately, a deeper understanding of scientific concept formation may also inform scientific pedagogy—not only by helping people learn existing scientific concepts, but by equipping them with the tools needed to discover and create new ones themselves.

\section{Code availability}

Source code, experiment scripts, and notebooks are available at: \url{https://github.com/omalagui/einstein_program_synthesis}

\section{Acknowledgements}

We are grateful to Stefano Profumo and Vidyesh Rao for helpful feedback. This research was supported by the Foundational Questions Institute and by the Faggin Presidential Chair Fund.

% \section*{Declarations}

% Some journals require declarations to be submitted in a standardised format. Please check the Instructions for Authors of the journal to which you are submitting to see if you need to complete this section. If yes, your manuscript must contain the following sections under the heading `Declarations':

% \begin{itemize}
% \item Conflict of interest/Competing interests (check journal-specific guidelines for which heading to use)
% \item Ethics approval and consent to participate
% \item Consent for publication
% \item Materials availability
% \end{itemize}

% \noindent
% If any of the sections are not relevant to your manuscript, please include the heading and write `Not applicable' for that section. 


\bibliography{sn-bibliography}% common bib file

\end{document}